\documentstyle[prl,aps,amssymb,graphicx]{revtex}
\newcommand{\beq}{\begin{equation}}
\newcommand{\eneq}{\end{equation}}
\begin{document}

\tolerance 10000

\twocolumn[\hsize\textwidth\columnwidth\hsize\csname %
@twocolumnfalse\endcsname

\draft

\title{Coordinate Representation of the Two-Spinon wavefunction and Spinon
Interaction}

\author {B. A. Bernevig$~^*$, D. Giuliano$~^{*\dagger}$ and R. B. Laughlin
$~^*$}

\address{$~^*$Department of Physics, Stanford University,
        Stanford, California 94305\\
	$~^\dagger$ Istituto Nazionale di Fisica della Materia (INFM),
         Unit\`a di Napoli, Napoli, Italy. }

\date{\today}
\maketitle
\widetext

\begin{abstract}
\begin{center}
\parbox{14cm}{ By deriving and studying the coordinate representation for 
the two-spinon wavefunction, we show that spinon excitations in the 
Haldane-Shastry model interact. The interaction is given by a short-range 
attraction and causes a resonant enhancement in the two-spinon wavefunction 
at short separations between the spinons. We express the spin susceptibility 
for a finite lattice in terms of the resonant enhancement, given by the
two-spinon wavefunction at zero separation. In the thermodynamic limit, the 
spinon attraction turns into the square-root divergence in the dynamical 
spin susceptibility.}

\end{center}
\end{abstract}

\pacs{
\hspace{1.9cm}
}
]

\narrowtext

\section{Introduction}

One of the most important issues in contemporary physics is spin 
fractionalization, which takes place in strongly-interacting 
one dimensional (1D) antiferromagnets. At odds with what one
would expect, the elementary excitation of these systems is not a spin-1
spin wave, but a gapless spin-1/2 excitation: the spinon.
 
Since their discovery within the framework of the Heisenberg model (HM),
by Fadeev and Takhtajan \cite{fadeev}, spinons are
believed to be noninteracting particles since, according to the Bethe-Ansatz
solution of the HM \cite{bethe}, the energy of a many-spinon solution is 
apparently given by the sum of the energies of each isolated spinon. In this
paper we challenge the non-interacting spinons idea through  a 
careful analysis of the  two-spinon dynamics in an exact solution, which shows 
 that they do actually  interact. 

 Extensive amount of theoretical studies proved that spin fractionalization is 
a generic phenomenon in one-dimensional spin-1/2 interacting antiferromagnets
\cite{conjec}. Indeed, the large-scale physics of all these systems is 
always the same, given by ``spinon gas'' dynamics \cite{spinongas}.
Therefore, without any loss of generality, one can choose to analyze a
 particular model, 
where the excitations are easier  to visualize.

The simplest model that describes the properties of spinons is the 
Haldane-Shastry model (HSM) \cite{haldane,shastry}, where  spins-1/2
located at the sites of a circular lattice antiferromagnetically 
interact and the interaction is inversely proportional to the 
square of the chord between the two sites.  
In this paper we investigate 
the basic features of the HSM by employing a
formalism based on analytic variables on the unit
radius circle \cite{chia}. 
By using real space coordinates, 
the spin-1/2 excitations become easier to construct and visualize than 
by making use of plane waves \cite{hald2,hald3}.  The formalism can 
be easily generalized to the study of other models. Such a formalism allows 
us to write a ``real space'' representation of the two-spinon wavefunction.

By analyzing the real space two-spinon wavefunction, we
show that spinons scatter by means of a short-ranged attractive potential and 
analyze in detail the physical consequences of the existence of this 
potential. The short-rangeness of the
interaction makes spinons free when they are widely separated. 
However, from the exact solution of the Schr\"odinger equation for two 
spinons we find that the
amplitude of the wavefunction is greatly enhanced when they are on
top of each other, phenomenon which we refer to as ``resonant
enhancement". While the density of states is
uniform at low energy, resonant enhancement causes the overlap between
the wavefunction for the localized spin wave and that for the spinon
pair to be significant, but not enough to create a two-spinon bound state. 
The corresponding matrix element is enhanced so 
as to make the spin-1 excitation absolutely unstable.

Physical consequences of the instability of the spin-wave appear in the 
functional form of the dynamical spin susceptibility (DSS), $\chi_q (\omega)$.
 The DSS is the 
Fourier transform of the spin-spin correlation function.
Its functional form can be experimentally tested
by means, for instance, of neutron scattering experiments, the probed 
quantity being the spectral density of states, $1/\pi
{\rm Im} \chi_q ( \omega)$  \cite{tennant}. A system with a stable
spin-1 excitation would show a sharp pole in ${\rm Im} \chi_q ( \omega)$
at the corresponding dispersion relation, $\omega = \omega ( q )$. On the
other hand, absolute instability of the spin wave against decay into spinons
 will generate a branch cut in ${\rm Im} \chi ( q , \omega)$ at the
threshold energy for the creation of a spinon pair, which is a 
signal of the opening of a decay channel, corresponding to the lack of 
spin wave integrity. Consequently, a sharp square-root singularity shows 
up at the threshold for the creation of a spinon pair, on top of the broadening
in the spectral density of states. Experiments performed onto quasi 
one-dimensional antiferromagnets provide clear evidence for broad spectra, 
while no sharp spin-1 resonance has been seen \cite{tennant}. 

An exact calculation of the DSS cannot, in general, be performed, even for
models exactly solvable with the Bethe-Ansatz. However, the HSM has the 
remarkable property that the wavefunction
for a spin-1 excitation is fully decomposed in the basis of the two-spinon
eigenstates \cite{hazi}. This allows us to write an exact expression for
the DSS even for a finite lattice, thus letting us explicitly show the 
relationship between the resonant enhancement (Fig.(\ref{fig5})) 
and the DSS.

The paper is organized as follows:
In section II we shortly review the HS Hamiltonian and its exceptional
symmetry; In Section III we introduce the ground state of the HSM and its
representation as a function of analytic variables on the unit circle. In terms
of the analytic variables the
ground state takes the same functional form as the fractional quantum Hall
wavefunction, which corresponds to a nondegenerate disordered spin 
singlet. We discuss at length several properties of the ground state,
how to derive the corresponding energy and the meaning of the 
disorder in the ground state;
In section IV we analyze the one-spinon solution and derive its relevant
properties; In section V we focus onto the two spinon solution. We derive
the energy eigenvalues, the corresponding eigenvectors, and their norm.
 A discussion about spinon statistics is provided at the end of the Section;
Sections VI and VII contain the key results of our work; In Section VI we 
write the Schr\"odinger equation for the two-spinon wavefunction, whose 
solutions are hypergeometric polynomials. From the behavior of the two-spinon 
wavefunction, we infer the nature of the interaction
between spinons: a short-range attraction.  The physical 
consequences of  such an interaction
are discussed at length in Section VII, where we derive an exact closed-form
expression for the dynamical Spin Susceptibility in terms of the two-spinon
wavefunctions and rigorously prove that the DSS is fully determined by spinon
interaction. In the thermodynamic limit spinon interaction turns into the
square root divergence in the DSS;
In Section VIII we provide our main conclusions.

\section{Haldane-Shastry Hamiltonian}

The Haldane-Shastry
 model \cite{haldane,shastry} is defined on a lattice with periodic
boundary conditions. Let $N$ be the number of sites.
Let $z_\alpha$, with $z_\alpha^N=1$, be a complex number representing 
a lattice site on which a spin-1/2 electron resides, and let  $\vec{S}_\alpha$
 be a Heisenberg spin operator acting on that electron. The Haldane-Shastry 
Hamiltonian takes the form:

\begin{equation}
{\cal H}_{HS} = J \left( \frac{2 \pi}{N} \right)^2
\sum_{\alpha < \beta}^N \frac{\vec{S}_\alpha \cdot
\vec{S}_\beta}{ | z_\alpha - z_\beta |^2 } \;\;\; ,  
\label{hsh}
\end{equation}

\noindent
where $J$ is the coupling strength. 
The interaction is an analytic function of the coordinates. This is related 
to the property of a complex variable $z$ laying on the unit circle: 
$z^* = z^{-1}$ which implies:

\begin{displaymath}
\frac{1}{ | z_\alpha - z_\beta |^2} = - \frac{ z_\alpha z_\beta}{ ( z_\alpha -
z_\beta )^2} \;\;\; .
\end{displaymath}
\noindent
The representation in terms of the analytic variables $z_\alpha$, which we
will use throughout the paper, comes out to be very useful for describing 
the properties of spinons in real space.
The Hamiltonian in Eq.(\ref{hsh}) is clearly invariant under spin rotations
generated by the total spin:

\begin{equation}
[ {\cal H}_{HS} , \vec{S} ] = 0
\; \; \; \; \; \; \; \; \;
\vec{S} = \sum_\alpha^N \vec{S}_\alpha \;\;\; .
\end{equation}

\noindent
It also possesses an additional symmetry generated by a vector operator
independent on $\vec{S}$:

\begin{equation}
[ {\cal H}_{HS} , \vec{\Lambda} ] = 0
\; \; \; \; \; 
\vec{\Lambda} = \frac{i}{2} \sum_{\alpha \neq \beta}
\biggl( \frac{z_\alpha + z_\beta}{z_\alpha - z_\beta} \biggr)
(\vec{S}_\alpha \times \vec{S}_\beta) \;\; .
\label{lambda}
\end{equation}

\noindent
That $\vec{\Lambda}$ commutes with ${\cal H}_{HS}$ can be seen as follows:

\[
[ {\cal H}_{HS} , \vec{\Lambda} ] 
\]

\begin{displaymath}
= \sum_{j \neq k} \sum_{\alpha \neq \beta} \frac{z_{j} + z_{k}}
{z_{j} - z_{k}} \frac{1}{\mid \! z_{\alpha} - z_{\beta} \! \mid^2}
\; [ (\vec{S}_{j} \times \vec{S}_{k}) , (\vec{S}_{\alpha} \cdot
\vec{S}_{\beta})]
\end{displaymath}

\begin{displaymath}
= 4 i \sum_{j \neq k \neq \ell}
\frac{z_{j} + z_{k}}{z_{j} - z_{k}} \frac{1}{ \mid \! z_{j} -
z_{\ell} \! \mid^2 } \biggl[ (\vec{S}_{j} \cdot \vec{S}_{k}) \;
\vec{S}_{\ell} - (\vec{S}_{\ell} \cdot \vec{S}_{k}) \; \vec{S}_{j}
\biggr]
\end{displaymath}

\begin{equation}
+ i \sum_{j \neq k} \frac{z_{j} + z_{k}}{z_{j} - z_{k}}
\frac{1}{ \mid \! z_{j} - z_{k} \! \mid^2 } \; (\vec{S}_{j} -
\vec{S}_{k} ) = 0  \;\;\;  .
\end{equation}

\noindent
Although they both commute with ${\cal H}_{HS}$, $\vec{S}$ and $\vec{\Lambda}$
do not commute between themselves, being that $\vec{\Lambda}$ is a vector, as
shown by the commutation relations:

\begin{equation}
[ S^a , S^b ] = i \; \epsilon^{abc} S^c
\; \; \; \; \; \; \; \; \; \; \; \; \;
[ S^a , \Lambda^b ] = i \; \epsilon^{abc} \Lambda^c
\; \; \; ,
\label{vector}
\end{equation}

\noindent
From the commutators in Eq.(\ref{vector}) it follows
that ${\cal H}_{HS}$, $S^2$, and $(\vec{\Lambda} \cdot \vec{S})$
all commute with each other.
The extra symmetry of ${\cal H}_{HS}$ is the reason for the exceptional
degeneracy of the energy eigenstates, as pointed out and discussed in
\cite{haldane}. The algebra generated by the two vector symmetries of
${\cal H}_{HS}$ is referred to as Yangian and is discussed in 
\cite{hald2,hald3}.
$\vec{\Lambda}$ can be physically interpreted as the spin 
current operator for the HSM, as we show in Appendix C.

Starting from next section we will review the properties of the ground state
and of the one- and two-spinon excited states of the HSM. This will allow us to
to define the formalism we will use in order to describe the relevant 
physical properties of the model.

\section{Ground State}

Let  $N$ be even. We proceed by first giving the representation 
of the ground state $ | \Psi_{GS} \rangle$ in terms of the $z$-coordinates and 
then proving that it is the actual ground state of ${\cal H}_{HS}$. 
$ | \Psi_{GS} \rangle $ is defined in terms of its projection onto the set of states
with $M=N/2$ spins up and the remaining spins down. If $z_1 , \ldots , z_{M}$
are the coordinates of the up spins, one defines the state $
 | z_1 , \ldots , z_{M} \rangle$ as: $ | z_1 , \ldots , z_{M} \rangle = 
\prod_{j=1}^M S_j^+
\prod_{\alpha = 1}^N c_{\alpha \downarrow}^\dagger | 0 \rangle$ where 
$ | 0 \rangle$ is the empty state. The projections are given by:

\begin{equation}
\Psi_{GS} (z_1, ... , z_{M}) = 
\prod_{j<k}^{M} (z_j - z_k)^2
\prod_{j=1}^{M} z_j  \;\;\; ,
\label{gstate}
\end{equation}

\noindent
where  $z_1 , ... , z_{M}$ denote the locations of the  $\uparrow$ 
sites all others being  $\downarrow$. We can imagine the spin system as 
a 1-dimensional string of boxes populated by hard-core bosons, the 
$\downarrow$ spin state
corresponding to an empty box and the $\uparrow$ spin state corresponding
to an occupied one. The total number of bosons is conserved, as it 
is physically the same thing as the eigenvalue of $S^z$. Let us, now, 
review the main properties of $\Psi_{GS} ( z_1 , \ldots , z_{M} )$.

\subsection{The norm of $\Psi_{GS}$}

$\Psi_{GS} ( z_1 , \ldots , z_{M} )$ is a homogeneous polynomial of degree
$N-1$ in the variables $z_1 , \ldots , z_{M}$.  
Its norm can be computed by using the following identity:

\[
C_M = \sum_{ z_1 , \ldots , z_{M}} \prod_{i<j}^M
| z_i - z_j |^4
\]

\beq
= ( \frac{N}{2 \pi i })^M \oint \frac{ d z_1}{  z_1} \ldots \oint 
\frac{ d z_{M}}{  z_{M} } \prod_{ i \neq j }^M
 ( 1 - \frac{ z_i }{ z_j} )^2 \;\;\; , 
\label{wili}
\eneq
\noindent
where the integrals are calculated on the circle of radius 1. 
The integral in Eq.(\ref{wili}) has been evaluated by Wilson \cite{kwil}.
The result is:

\beq 
( \frac{1}{ 2 \pi i })^M
\oint \frac{ d z_1}{  z_1} \ldots \oint 
\frac{ d z_M}{  z_M } \prod_{ i \neq j }^M ( 1 - \frac{ z_i
}{ z_j} )^2 = \frac{ ( 2 M )!}{ 2^M} \; ,
\label{wilson2}
\eneq
\noindent
therefore:

\beq
C_M= \frac{ ( 2 M )!}{ 2^M} N^M \;\;\; .
\label{wili31}
\eneq

\subsection{Singlet Sum Rule}

We shall prove that the ground state is a spin singlet by showing that  
$| \Psi_{GS} \rangle $ is annihilated by both $S^z$ and $S^-$.
$S^z|\Psi_{GS}\rangle = 0$ because $|\Psi_{GS} \rangle$ has an equal 
number of $\uparrow$ and $\downarrow$ spins

      \begin{displaymath}
      [S^{-}\Psi_{GS}](z_{2}, \ldots ,z_{M}) = \sum_{\alpha = 1}^{N}
      \langle z_2 , \ldots , z_{M} | S_\alpha^- | \Psi_{GS} \rangle 
      \end{displaymath}

      \begin{equation}
	=
      \lim_{z_1 \rightarrow 0} \;
      \sum_{\ell = 1}^{N-1} \frac{1}{\ell !}\biggl\{
      \sum_{\alpha = 1}^{N} z_{\alpha}^{\ell} \biggr\}
      \frac{\partial^{\ell}}{\partial z_{1}^{\ell}} \Psi_{GS}
      (z_{1}, \ldots ,z_{M})  = 0
      \label{singlet}
      \end{equation}

\noindent
      since  $\sum_{\alpha = 1}^{N} z_{\alpha}^{\ell} = N \; \delta_{\ell 0}
      \pmod{N} \; \; \; .$

\noindent

\subsection{Coordinate Invariance}

Spin rotational invariance implies that $\Psi_{GS}$ is invariant under interchange of $\uparrow$ and $\downarrow$ coordinates.
More generally, the quantization axis can be taken to be an arbitrary
direction in spin space.  Denoting the
sites complementary to $z_1 , ..., z_{M}$ by $\eta_1 , ..., \eta_{M}$,
so that

\begin{equation}
\prod_k^{M} (z - z_k )(z - \eta_k) = z^{N} - 1 \; \; \; ,
\end{equation}

\noindent
we have for fixed $j$

\begin{equation}
\prod_{ k \neq j}^M (z_j - z_k) (z_j - \eta_k ) =
\lim_{z \rightarrow z_j} \frac{z^{N} - 1}{z - z_j} = N \; z_j^{N-1}
\; \; \; ,
\end{equation}

\noindent
and thus

\begin{displaymath}
\prod_{j < k}^M (z_j - z_k )^2 \prod_j^M z_j
= N \prod_{j , k}^M \frac{1}{z_j - \eta_k}
\end{displaymath}

\begin{equation}
= (-1)^{M} \prod_{j < k}^M (\eta_j - \eta_k )^2 \prod_j^M \eta_j
\; \; \; .
\end{equation}

\subsection{Reality}

The ground state is its own complex conjugate, and therefore is a real number:

      \begin{displaymath}
      \Psi_{GS}^{*}(z_{1} , \ldots , z_{M}) = \prod_{j < k}^{M}
      (z^{*}_{j} - z^{*}_{k})^2 \prod_{j}^{M} z^{*}_{j}
      \end{displaymath}

      \begin{equation}
      = \prod_{j < k}^{M}
      (z_{k} - z_{j})^2 \prod_{j}^{M} z_{j}^{1-N} =
      \Psi_{GS} (z_{1} , \ldots , z_{M}) \;\;\; .
      \end{equation}

\subsection{Translational Invariance}

The crystal momentum of the state, $q$, is defined (mod $2 \pi$) by the
equation:

\begin{equation}
\Psi_{GS} ( z_1 z , \ldots , z_{M} z ) = e^{ iq}
\Psi_{GS} ( z_1 , \ldots , z_{M}  ) \;\;\;  ,
\label{cm}
\end{equation}

\noindent
where $z$=$\exp (2 \pi i  /N)$ . From Eq.(\ref{cm}) it comes out that 
$q$ can be either 0 or $\pi$, according to whether $N$ is divisible by four 
or not. $\Psi_{GS}$ equals itself, up to an overall minus sign, when translated
by one lattice constant.

\subsection{Disordered State}

 $| \Psi_{GS} \rangle$ is a disordered state. The way spin-spin correlations 
fall off with the distance defines whether a state of a magnetic system 
takes order or not. The relevant quantity is the spin-spin correlation
function, $\chi (z_\alpha) =
\langle \Psi_{GS} | S_0^+ S_\alpha^- | \Psi_{GS} \rangle /
\langle \Psi_{GS} | \Psi_{GS} \rangle$, which can be expressed in terms
of two-spinon wavefunctions only, as we show in Section VII.

\begin{figure}
\includegraphics*[width=0.87\linewidth]{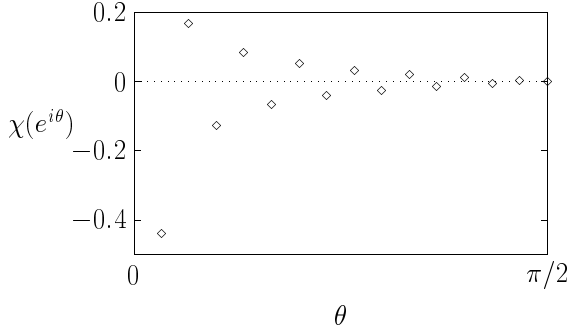}
\caption{Spin-spin correlation function decay for $N=60$.}
\label{fig2}
\end{figure}

\noindent
One-dimensional systems do not break continuous symmetries, 
so they  are not alleged to  order. However, there is a substantial difference 
between half-integer spin chains and integer spin ones \cite{conjec}. 
Both have a disordered ground state, but the former have excitations 
above the ground  state which are gapless in the thermodynamic limit 
while the latter have a gap that survives the thermodynamic
limit and is given by $\Delta = \hbar v / \xi$, where $v$ is the spin-wave
velocity of a nearby ordered state and $\xi$ is the correlation length. The
consequence of this is that the fall-off of the spin correlations in the
ground state of half-odd spin chains is not as abrupt as for integer-spin
chains, where the correlations are suppressed within one or two lattice
spacings (``Haldane's conjecture''). Fig.(\ref{fig2}) shows that the 
behavior of the HSM is the one  expected for half-odd spin chains. 
Correlations decay as $(-1)^x/x$, according to Haldane's conjecture.

\subsection{Ground State Energy}
 $| \Psi_{GS} \rangle$ is an eigenstate of ${\cal H}_{HS}$ with eigenvalue:

\begin{equation}
{\cal H}_{HS} | \Psi_{GS} \rangle  = -J \left( \frac{\pi^2}{24} \right)
\left(N + \frac{5}{N} \right) | \Psi_{GS} \rangle  \; \; \; .
\end{equation}

\noindent
We trade sums over spins on the
lattice for derivative operators that are understood to act onto the
analytic extension of $\Psi_{GS} ( z_1 , \ldots , z_{M} )$, in which
the $z_j$'s are allowed to take any value on the unit circle. After
computing the derivatives, we constrain them again to lattice
sites.
We begin by observing that $[S_\alpha^+ S_\beta^- \Psi_{GS} ]
(z_1 , ... , z_{M})$ is identically zero unless one of the arguments
$z_1 , ... , z_{M}$ equals $z_\alpha$. We have

\begin{displaymath}
[ \biggl\{ \sum_{\beta \neq \alpha}^{N} \frac{ S_{\alpha}^{+}
S_{\beta}^{-}}{\mid \! z_{\alpha} - z_{\beta} \! \mid^2 }
\biggr\} \Psi_{GS}] (z_{1}, \ldots , z_{M})
\end{displaymath}

\begin{displaymath}
= \sum_{j=1}^{M} \sum_{\beta \neq j}^{N} \frac{1}{\mid \! z_j
- z_{\beta} \! \mid^2 } \Psi_{GS} (z_{1}, \ldots , z_{j-1}, z_{\beta} ,
z_{j+1}, \ldots , z_{M})
\end{displaymath}

\begin{displaymath}
= \sum_{j=1}^{M} \sum_{\ell = 0}^{N-2} \biggl\{
\sum_{\beta \neq j}^{N} \frac{ z_{\beta} (z_{\beta} - z_{j})^{\ell}}
{ \ell ! \mid \! z_{j} - z_{\beta} \! \mid^2} \biggr\}
(\frac{\partial}{\partial z_{j}})^{\ell} \biggl\{ \frac{ \Psi_{GS}
(z_{1}, \ldots , z_{M})}{z_{j}} \biggr\}
\end{displaymath}

\begin{equation}
=  \sum_{\ell=0}^{ N -2} \sum_{j = 1}^M \frac{ z_j^{ \ell +1}}{\ell !} A_l
( \frac{ \partial^\ell}{ \partial z_j^\ell} ) \biggl\{
\frac{ \Psi_{GS} ( z_1 , \ldots , z_{M} )}{  z_j} \biggr\} \;\;\; .
\label{engs1}
\end{equation}

\noindent
The coefficients $A_l$ are evaluated in appendix B. Their
remarkable property is that they are zero for $N>l>2$. Hence, 
Eq.(\ref{engs1}) can be rewritten as:

\begin{displaymath}
 \sum_{j=1}^{M} \biggl\{ \frac{(N-1)(N-5)}{12} z_{j} - \frac{N-3}{2}
 z_{j}^2 \frac{\partial}{\partial z_{j}}   
\end{displaymath}

\begin{displaymath}
+ \frac{1}{2} z_{j}^3
 \frac{\partial^2} {\partial z_{j}^2} \biggr\}
  \biggl\{ \frac{ \Psi_{GS} (z_{1},
 \ldots , z_{M})}{ z_{j}} \biggr\}
\end{displaymath}

\begin{displaymath}
= \biggl\{ \frac{N(N-1)(N-5)}{24} - \frac{N-3}{2}
\sum_{j \neq k}^{M} \frac{2 z_{j}}{z_{j} - z_{k}}
\end{displaymath}

\begin{displaymath}
+ \sum_{j \neq k \neq m}^{M} \frac{2 z_{j}^2}
{(z_{j} - z_{k})(z_{j} - z_{m})}+ \sum_{j \neq k}^{M} \frac{ z_{j}^2}
{(z_{j} - z_{k})^2} \biggr\} 
\end{displaymath}

\begin{displaymath}
\times  \Psi_{GS} (z_{1}, \ldots ,z_{M}) 
\end{displaymath}

\begin{equation}
=
\biggl\{ - \frac{N}{8} - \sum_{j \neq k}^{M}
\frac{1}{\mid \! z_{j} - z_{k} \! \mid^2 } \biggr\}
\Psi_{GS} (z_{1}, \ldots , z_{M}) \; \; \; .
\label{kings}
\end{equation}

\noindent
In Eq.(\ref{kings}) we have made use of the rule:

\begin{displaymath}
\frac{z_{\alpha}^2}{(z_{\alpha}-z_{\beta})(z_{\alpha}-z_{\gamma})} +
\frac{z_{\beta}^2}{(z_{\beta}-z_{\alpha})(z_{\beta}-z_{\gamma})} 
\end{displaymath}

\begin{equation}
+ \frac{z_{\gamma}^2}{(z_{\gamma}-z_{\alpha})(z_{\gamma}-z_{\beta})} =
1 \;\;\; .
\end{equation}

\noindent
We also have:

\begin{displaymath}
[ \biggl\{ \sum_{\beta \neq \alpha}^{N} \frac{ S_{\alpha}^{z}
S_{\beta}^{z}}{\mid \! z_{\alpha} - z_{\beta} \! \mid^2 }
\biggr\} \Psi_{GS}] (z_{1}, \ldots , z_{M}) 
\end{displaymath}

\[
=
 \biggl\{ - \frac{N(N^2 - 1)}{48}
+ \sum_{j \neq k}^{M} \frac{1}{\mid \! z_{j} - z_{k} \!
\mid^2 } \biggr\} 
\]

\beq
\times
\Psi_{GS} (z_{1}, \ldots , z_{M})
\; \; \; .
\end{equation}

\noindent
This completes the proof, since

\[
{\cal H}_{HS} = \frac{J}{2}  (\frac{2\pi}{N})^2 \biggl\{ \sum_{\alpha
\neq \beta}^N \frac{ S^{+}_{\alpha} S^{-}_{\beta}}{ \mid \! z_{\alpha} -
z_{\beta} \! \mid^2 } 
\]

\beq
+ \sum_{\alpha \neq \beta}^N \frac{ S^{z}_{\alpha}
S^{z}_{\beta}}{ \mid \! z_{\alpha} - z_{\beta} \! \mid^2 } \biggr\} \;\;\; .
\label{gsen}
\end{equation}

\noindent
The wavefunction $\Psi_{GS} ( z_1 , \ldots , z_{M} )$ was first introduced
by Haldane and Shastry  \cite{haldane,shastry} in analogy to the exact 
Sutherland solution of the continuum limit of the problem \cite{sutherland}. 
The proof that this wavefunction is 
the actual ground state of ${ \cal H}_{HS}$ is a consequence of the 
factorization of the HS Hamiltonian, as we are going to discuss next.

\subsection{Factorization of ${ \cal H}_{HS}$}

In Appendix D we prove that ${ \cal H}_{HS}$ can be written as:

\begin{displaymath}
{ \cal H}_{HS} = J ( \frac{ 2\pi}{ N} )^2 \biggl[ \frac{2}{9} \sum_\alpha^N 
\vec{D}_\alpha^\dagger \cdot \vec{D}_\alpha 
\end{displaymath}

\begin{equation}
- \frac{ N ( N^2 + 5 )}{ 48} + \frac{ N +1}{ 12} \vec{S}^2 \biggr] \;\;\; .
\label{factri}
\end{equation}

\noindent
The operators $D_\alpha$ are given by:

\begin{equation}
\vec{D}_\alpha
= \frac{1}{2} \sum_{\beta \neq \alpha}^N \frac{z_\alpha +
z_\beta} {z_\alpha - z_\beta} \biggl[ i ( \vec{S}_\alpha \times
\vec{S}_\beta ) + \vec{S}_\beta \biggr] \; \; \; .
\end{equation}

\noindent
and they annihilate $ | \Psi_{GS} \rangle$ (see Appendix D). Eq.(\ref{factri}) 
implies that $| \Psi_{GS} \rangle$ is the ground state of the HSM because
  ${\cal H}_{HS}$ 
can be written as a constant plus nonnegative definite operators, and 
the only state satisfying the requirement of minimum energy is $ 
| \Psi_{GS} \rangle$.

\subsection{Degeneracy}

The HSM ground state is not degenerate, but is nearly so. We already
pointed out that half-odd spin magnets have a gapless spectrum. In the
next sections we will see that elementary excitations above the ground
state are spinons and that their spectrum is relativistic.
In particular, at the endpoints of the Brillouin zone, the energy becomes 
the same as the ground state energy, modulo corrections that are sub-leading in
the thermodynamic limit. This means that, in principle, one can have many
states with the same energy as the ground state which are distinguished from 
one another by their number of spinons.

An example is provided by the singlet state of two spinons
with total momentum $\pi$. It is given by:

      \begin{equation}
      \Psi_S (z_1, ... , z_{M}) =
      \prod_{j<k}^{M} (z_j - z_k)^2
      \biggl[ 1 - \prod_{j=1}^{M} z_j^2 \biggr] \; \; \; .
      \label{alternate}
      \end{equation}

\noindent
Its energy is given by:

      \begin{equation}
      {\cal H}_{HS} | \Psi_{S}  \rangle = -J \left( \frac{\pi^2}{24} \right)
      \left(N - \frac{7}{N} \right) | \Psi_S \rangle \; \; \; .
      \end{equation}

\noindent
and is the energy of the ground state plus corrections that
go to zero in the thermodynamic limit.

\subsection{Spin current}

We now show that $ | \Psi_{ GS} \rangle$ is an eigenstate of $\Lambda^z$ 
belonging to the 0-eigenvalue.
The action of $\Lambda^z$ on the ground state gives:

\begin{equation}
\Lambda^z | \Psi_{GS } \rangle = 0  \;\;\; .
\label{lgs}
\end{equation}

\noindent
Eq.(\ref{lgs}) can be proved as follows:

\begin{displaymath}
[ \Lambda^z \Psi_{GS} ] ( z_1 , \ldots , z_{M} )
\end{displaymath}

\begin{displaymath}
= \frac{1}{2} \sum_{ j = 1}^M \sum_{ \beta \neq j}^N
z_\beta \left( \frac{ z_j + z_\beta }{  z_j - z_\beta} \right)  
\end{displaymath}

\begin{displaymath}
\times \sum_{ l = 0}^{ N- 1}
\frac{ (z_j - z_\beta)^l}{ l!} \frac{ \partial}{ \partial z_j^l}
\biggl\{ \frac{ \Psi_{GS} ( z_1 , \ldots , z_{M} ) }{ z_j} \biggr\}
\end{displaymath}

\[
= \left[ - \frac{N}{4} ( N- 2) + \frac{N}{ 4} ( N - 2 ) \right]
\]

\beq
\times
 \Psi_{GS} ( z_1 , \ldots , z_{M} ) = 0 \;\; ,
\label{lgss}
\end{equation}
\noindent
where we have made use of the results of Appendix B and of the technique
described in detail in subsection III.F.

\section{One-spinon wavefunction.}

At odds with the naive idea that the elementary excitations for 
interacting magnets are integer spin states (spin flips), Faddev and 
Takhtajan \cite{fadeev} first conjectured that one-dimensional half odd
spin chains exhibit excitations given by spinons carrying half-odd spin. 
For a chain with an even number of sites, the ground state is a
disordered spin singlet but, if the number of sites is odd, the
minimum possible value for the total spin is 1/2. In the thermodynamic 
limit, it  makes no difference whether one begins with an odd or 
an even number of sites. The short-rangeness of the correlations in the 
ground state makes it insensitive to the boundary conditions, so, in the 
thermodynamic limit, there is no  way to distinguish between chains  
with odd or even number of sites. States with half-odd spin are, then, 
alleged to appear as eigenstates of ${\cal H}_{HS}$ with an odd number of 
spinons. In this section we shall present the one-spinon wavefunction and
discuss its properties. 
Following Haldane \cite{haldane} we consider a wavefunction of the
general form

\begin{equation}
\Psi (z_{1} , \ldots , z_{M}) = \Phi (z_{1}, \ldots ,z_{M}) 
\prod_{j < k}^{M} (z_{j} - z_{k})^2 \prod_{j}^{M} z_{j} 
\label{mspin}
\end{equation}

\noindent
where $z_1, \ldots , z_M$ denote the position of the up spins. Here 
 $\Phi$ is a homogeneous symmetric polynomial of
degree less than $N - 2M + 2$ in each variable.  This latter condition 
causes $\Psi$ to be a polynomial of degree less than $N+1$ in each of 
its variables $z_{j}$, and thus allows the Taylor expansion technique used 
for the ground state to be applied.  Doing so, we find that

\begin{displaymath}
{\cal H}_{HS} \Psi = \frac{J}{2} (\frac{2\pi}{N})^2
\biggl\{ \lambda +\frac{N}{48}(N^2 - 1)
\end{displaymath}

\begin{equation}
 + \frac{M}{6}
(4 M^2 - 1) - \frac{N}{2} M^2 \biggr\} \; \Psi \; \;\; ,
\label{diffs}
\end{equation}

\noindent
provided that $\Phi$ satisfies the eigenvalue equation for $\lambda$

\begin{displaymath}
 \frac{1}{2} \biggl\{ \sum_{j}^{M} z_{j}^2 \frac{\partial^2
\Phi} {\partial z_{j}^2} + \sum_{j \neq k}^M \frac{4
z_{j}^2} {z_{j}-z_{k}} \frac{\partial \Phi}{\partial z_{j}}
\biggr\}
\end{displaymath}

\begin{equation}
- \frac{N-3}{2} \sum_{j}^{M} z_{j} \frac{\partial \Phi}
{\partial z_{j}} = \lambda \Phi \; \; \; .
\label{eig2}
\end{equation}

\noindent

\subsection{One-Spinon Spin Doublet}

We look for one-spinon and two-spinon wavefunctions in the functional
form given by Eq.(\ref{mspin}). Here we analyze the one-spinon wavefunction.

\begin{figure}
\includegraphics*[width=0.87\linewidth]{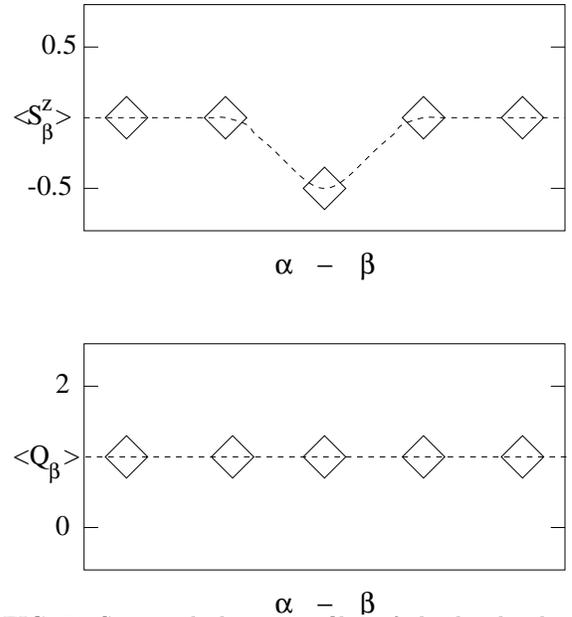}
\caption{Spin and charge profiles of the localized spinon
         $|\Psi_\alpha \rangle$ defined by Eq.(\ref{spinon1}).  
         The dotted lines are a guide to the eye.}
\label{fig3}
\end{figure}

\noindent
Let the number of sites $N$ be odd and let

\[
\Psi_{\alpha} (z_{1} , ... , z_{M}) 
\]

\beq
=
\prod_{j}^{M} (z_{\alpha} - z_{j}) \;
\prod_{j < k}^{M} (z_{j} - z_{k})^2
\prod_{j}^{M} z_{j} \; \; \; ,
\label{spinon1}
\end{equation}

\noindent
where $M = (N - 1)/2$. This is a $\downarrow$ spin on site $\alpha$
surrounded by an otherwise featureless singlet sea. 
It is worth stressing that Eq.(\ref{spinon1}) makes perfect sense for any
 $z_\alpha$ on the unit circle. Nevertheless, as $z_\alpha$ coincides with a 
lattice site, it represents a spin $\downarrow$ localized at the corresponding
site. The spin density of the corresponding state, plotted as a function of
the spinon position, will be uniformly zero, as appropriate for the
disordered spin singlet, except for an abrupt dip centered at $z=z_\alpha$
(see Fig.(\ref{fig3})). Such a dip is what we refer to as ``real space 
representation'' of a spinon at $z_\alpha$. Hence, a spinon can be visualized
as a local defect in an otherwise featureless singlet sea. This defect behaves
 like a real quantum mechanical particle, as we will show in the following.

By definition, $\Psi_\alpha$ is an eigenstate of $S^z$ with eigenvalue
-1/2.
In order to prove that it is a spin-1/2 state, we need to show
that $S^-$ annihilates it. Indeed, per Eq.(\ref{singlet}) we have:

\begin{equation}
\sum_{\beta \neq \alpha}^N S_{\beta}^{-} \Psi_{\alpha} = 0
\; \; \; ,
\label{singlet2}
\end{equation}
\noindent
which proves that $\Psi_\alpha$ is the spin-1/2 component of a spin 
doublet.

\subsection{One-Spinon Energy}

\noindent
Eq.(\ref{spinon1}) corresponds to a particular choice of $\Phi$ in
Eq.(\ref{mspin}), given by:

\begin{equation}
\Phi ( z_1, \ldots , z_M ) =
\Phi_\alpha ( z_1 , \ldots ,z_M ) = \prod_{j=1}^M ( z_\alpha - z_j ) \;\; .
\end{equation}

\noindent
Eq.(\ref{eig2}), once written for the state $\Phi_\alpha$, takes the form:

\begin{displaymath}
\biggl\{ M ( M - 1) - z_\alpha^2 \frac{ \partial^2}{ \partial z_\alpha^2}
- \frac{ N - 3}{2} \left[ M - z_\alpha \frac{ \partial}{ \partial z_\alpha }
\right] \biggr\} \Phi_\alpha
\end{displaymath}

\begin{equation}
= \lambda \Phi_\alpha \;\;\; .
\label{sdif2}
\end{equation}

\noindent
The eigenstate of ${\cal H}_{HS}$ is given by:

\begin{equation}
\Psi_m ( z_1 , \ldots , z_M ) = \frac{1}{N} \sum_{ \alpha = 1}^N 
 (z_{\alpha}^{*})^m
\Psi_\alpha ( z_1 , \ldots , z_M )
\end{equation}

\noindent
and the energy eigenvalue is

\[
{\cal H}_{HS} | \Psi_{m} \rangle
= \biggl\{ - J (\frac{\pi^2}{24}) ( N -
\frac{1}{N}) 
\]

\beq
+ \frac{J}{2} (\frac{2\pi}{N})^2 m(\frac{N-1}{2} - m)
\biggr\} | \Psi_{m} \rangle \;\;\; ,
\end{equation}

\noindent
with $ 0 \leq m \leq (N-1)/2$ and $\lambda = m ( ( N - 1 ) /2 - m)$.

\subsection{Crystal Momentum}

\begin{figure}
\includegraphics*[width=0.87\linewidth]{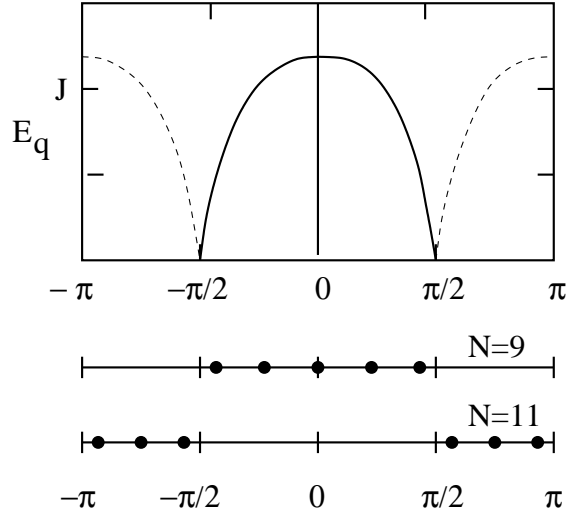}
\caption{Top: Spinon dispersion given by Eq.(\ref{dispersion}). Bottom: Allowed
        values of $q$ for adjacent odd $N$.}
\label{fig4}
\end{figure}

\noindent
The state $| \Psi_m \rangle $ is a propagating $\downarrow$ spinon with
crystal momentum

\begin{equation}
q = \frac{\pi}{2} N - \frac{2\pi}{N}(m + \frac{1}{4}) \pmod{2\pi}
\; \; \; ,
\label{momentum}
\end{equation}

\noindent
per the definition

\begin{equation}
\Psi_{m} (z_{1} z , \ldots , z_{M} z) = \exp (iq) \;
\Psi_{m} (z_{1} , \ldots , z_{M}) \; \; \; ,
\label{crys2}
\end{equation}

\noindent
where $z=\exp(2\pi i/N)$. Rewriting the eigenvalue as

\begin{equation}
{\cal H} | \Psi_{m} \rangle = \biggl\{ - J (\frac{\pi^2}{24}) ( N +
\frac{5}{N} - \frac{3}{N^2}) + E_{q} \biggr\} | \Psi_{m} \rangle 
\; \; ,
\end{equation}

\noindent
we obtain the dispersion relation

\begin{equation}
E ( q) = \frac{J}{2} \biggl[ (\frac{\pi}{2})^2 - q^2 \biggr] \pmod{\pi}
\label{dispersion} \;\;\; ,
\end{equation}

\noindent
plotted in Fig.(\ref{fig4}).  Note that the momenta available to the
spinon span only the inner or outer half of the Brillouin zone,
depending on whether $N - 1$ is divisible by 4 or not. The loss of half of
the states available for a regular fermion is a peculiar property of
spinon spectrum. No negative energy states appear, i.e., there is nothing
like an ``antispinon''. One can picture a spinon as either an electron or
a hole whose charge has been pulled out by the interaction. According to
such a picture, a spinon can arise either from an electron with the
same spin or from a hole with the opposite spin, which explains the halving
of the Brillouin zone.

The spinon dispersion at low energies is linear in $q$ with a velocity

\begin{equation}
v_{\rm spinon} = \frac{\pi}{2} J \; \; \; ,
\label{1dpen}
\end{equation}

\noindent
 The half-band of single elementary excitations for odd $N$ are the only 
$S=1/2$ states without extra degeneracies. The ground state of the 
odd-N spin chain is 4-fold degenerate and is
given by $| \Psi_m \rangle$ for $m = 0$ and $(N-1)/2$ and their $\uparrow$
counterparts.  This corresponds physically to a ``left-over'' spinon
with momentum $\pm \pi$.

\subsection{Spin-Current}

We now study the action of $\Lambda^z$ on the state for one propagating spinon.
Working as for the ground state one gets:

\begin{equation}
\Lambda^z | \Psi_m \rangle = \biggl\{ \frac{ N-1}{ 4} - m \biggr\}
| \Psi_m \rangle  \; \; \; ,
\label{l1sp}
\end{equation}
\noindent
and the eigenvalue of $\Lambda^z$ comes out to be proportional to the spinon 
velocity

\begin{equation}
\frac{d E (q) }{dq} = - \frac{2 \pi J}{N} \biggl\{ \frac{N-1}{4} - m
\biggr\} \;\;\; .
\end{equation}
\noindent
Eq.(\ref{l1sp}) is proven by first letting $\Lambda^z$ act on the
state $\Psi_\alpha$ defined in Eq.(\ref{spinon1}):

\begin{displaymath}
[ \Lambda^z  \Psi_\alpha ] ( z_1 , \ldots , z_M )
\end{displaymath}

\begin{displaymath}
= \frac{1}{2} \sum_{ j=1}^M \sum_{ \beta \neq j}^N
z_\beta \left( \frac{ z_j + z_\beta }{  z_j - z_\beta}
\right) \sum_{ l = 0}^{ N- 1}
\frac{ (z_j - z_\beta)^l}{ l!} 
\end{displaymath}

\begin{displaymath}
\times \frac{ \partial}{ \partial z_j^l}
\biggl\{\frac{ \Psi_{\alpha} ( z_1 , \ldots , z_M ) }{ z_j} \biggr\}
\end{displaymath}

\begin{displaymath}
= \frac{1}{2}
\biggl\{ - M ( N-2) + 2 \sum_{j =1}^M \sum_{i \neq j }^M
\frac{ z_j}{ z_j - z_i} 
\end{displaymath}

\beq
+ 2 \sum_{ j=1 }^M \frac{z_j}{ z_j - z_\alpha}
\biggr\} \Psi_\alpha ( z_1 , \ldots , z_M ) \;\;\; .
\label{los1}
\end{equation}
\noindent
After taking $M = (N - 1)/2$ one gets:

\begin{displaymath}
[ \Lambda^z  \Psi_\alpha ] ( z_1 , \ldots , z_M)
\end{displaymath}

\begin{equation}
 =
\biggl\{ \frac{N-1}{4} - z_\alpha \frac{\partial}{ \partial z_\alpha }
\biggr\} \Psi_\alpha ( z_1,  \ldots , z_M ) \;\;\; ,
\end{equation}
\noindent
which, on the  basis of the states  $ | \Psi_m \rangle$,
 gives the result quoted in Eq.(\ref{l1sp}).

\subsection{The Norm}

The squared norm of $\Psi_m$ is defined as:

\beq
\langle \Psi_m|\Psi_m \rangle= \sum_{ z_1 , \ldots , z_M} | 
\Psi_m ( z_1 , \ldots , z_M ) |^2  \;\;\; .
\label{nor1}
\eneq
\noindent
By means of a simple algebraic procedure, we generated a recursion relation
 between $\langle \Psi_m|\Psi_m \rangle$ and 
$\langle \Psi_{m-1}|\Psi_{m-1} \rangle$.  Such a procedure
can be straightforwardly extended to the norm of the multiple-spinon states.
We discuss it at length in Appendix E. 
The induction relation is:

\beq
\frac{\langle \Psi_m|\Psi_m \rangle}{\langle \Psi_{m-1}|\Psi_{m-1} \rangle} = \frac{ ( m - \frac{1}{2} ) ( M - m + 1 )}{ m ( M - m + \frac{1}{2} )} \;\;\; .
\label{norma1}
\eneq
\noindent
This recursively gives:

\beq
 \langle \Psi_m|\Psi_m \rangle  = 
\frac{ \Gamma [ M + 1 ] \Gamma [ m + \frac{1}{2} ] \Gamma [ M - m + \frac{1}{2
} ] }{ \Gamma [ \frac{1}{2} ] 
\Gamma [ M + \frac{1}{2} ] \Gamma [ m + 1 ] \Gamma [ M - m + 1 ] 
} C_M
\label{norma2}
\eneq
\noindent
where $C_M$ is the over-all constant we have introduced in Eq.(\ref{wili}).

\section{Two-spinon wavefunction}

Let us now focus on the two-spinons  state. Spinons maintain their 
integrity when many of them are present.
 This does not mean that spinons are noninteracting. 
They can be separated at large distances, therefore
being asymptotic states of the system. However, they also scatter strongly
off each other by means of a short-ranged attractive potential.
The interaction between spinons is not enough to create two-spinon bound
states, but it generates a peculiar ``piling-up'' of the relative 
wavefunction when the two spinons are on top of each other (Fig.(\ref{fig5})).
This is the reason for the huge decay amplitude for a spin wave into 
a pair of spinons, that is the spin fractionalization. 

In this section we derive the  two-spinon eigenstates, their norm and
the corresponding value of the spin-current.
Moreover, we show that the appropriate statistics they obey is neither 
fermionic nor bosonic. They are semions, i.e., particles with 1/2 
fractional statistics. 

\subsection{Two-spinon Energy}

Two $\downarrow$-spinons can be pictured as two $\downarrow$-spins within an
otherwise featureless disordered sea.  The state with two $\downarrow$-spinons
 centered at  $z_\alpha$ and $z_\beta$, respectively, is given 
by ($N$ is even and $M=N/2 - 1$):

\[
\Psi_{\alpha \beta} (z_{1} , \ldots , z_{M}) 
\]

\beq
= \prod_{j}^{M} (z_{\alpha} - z_{j}) (z_{\beta} - z_{j}) \;
\prod_{j < k}^{M} (z_{j} - z_{k})^2
\prod_{j}^{M} z_{j} \;\;\; .
\label{2spinon}
\end{equation}

\noindent
As for the one-spinon case, $z_\alpha$ and $z_\beta$ are not necessarily
lattice sites. If they are, Eq.(\ref{2spinon}) represents a pair of spinons
at $z_\alpha$, $z_\beta$.
To derive the eigenvalue equation, we start from a wavefunction in the 
form of Eq.(\ref{mspin}),  where we take the function $\Phi$ to be equal to:

\begin{equation}
\Phi_{\alpha \beta } = \prod_{j}^M ( z_\alpha - z_j ) ( z_\beta - z_j) \;\;\; .
\label{2sp1} 
\end{equation}

\noindent
Eq.(\ref{eig2}) can be rewritten for $\Phi_{\alpha \beta}$, yielding:

\begin{displaymath}
\frac{1}{2} \biggl\{ \sum_{j=1}^{M} z_{j}^2 \frac{\partial^2
\Phi_{\alpha \beta}} {\partial z_{j}^2} + \sum_{j \neq k}^M
\frac{4 z_{j}^2} {z_{j}-z_{k}} \frac{\partial \Phi_{\alpha \beta}}
{\partial z_{j}} \biggr\} - \frac{N-3}{2} \sum_{j=1}^{M} z_{j}
\frac{\partial \Phi_{\alpha \beta}}
{\partial z_{j}}
\end{displaymath}

\begin{displaymath}
= \biggl\{ - \frac{z_\alpha^2}{z_\alpha - z_\beta}
\frac{\partial}{\partial z_\alpha}
- \frac{z_\beta^2}{z_\beta - z_\alpha}
\frac{\partial}{\partial z_\beta}
- z_\alpha^{2} \frac{\partial^2}{\partial z_\alpha^2}
- z_\beta^{2} \frac{\partial^2}{\partial z_\beta^2}
\end{displaymath}

\begin{displaymath}
+ (\frac{N-3}{2}) \biggl[
z_\alpha \frac{\partial}{\partial z_\alpha}
+ z_\beta \frac{\partial}{\partial z_\beta} \biggr]
+ \biggl[2M^2 - M(N-2) \biggr] \biggr\} \Phi_{\alpha \beta} 
\end{displaymath}

\begin{equation}
= \lambda  \Phi_{\alpha \beta} \;\;\; .
\label{eig3}
\end{equation}

\noindent
Let us now define the states $\Psi_{mn}$ as follows:

\[
\Psi_{mn} ( z_1 , \ldots , z_M )
\]

\beq
 = \sum_{\alpha,\beta }^N \frac{ (z_\alpha^*
)^m}{N} \frac{ ( z_\beta^* )^n}{ N} \Psi_{ \alpha \beta } 
( z_1 , \ldots , z_M)
\label{herethepsi}
\end{equation}

\noindent
 A set of linearly independent states may be constructed
by taking only the $\Psi_{ mn}$ with $M \geq m \geq n \geq 0$, which shows the 
 overcompleteness of the set of states $\Psi_{mn}$. 
On such a set of states Eq.(\ref{eig3}) becomes:

\begin{displaymath}
\frac{1}{2} \biggl\{ \sum_{j=1}^{M} z_{j}^2 \frac{\partial^2
} {\partial z_{j}^2} + \sum_{j \neq k}^M \frac{4
z_{j}^2} {z_{j}-z_{k}} \frac{\partial }{\partial z_{j}}
- (N-3) \sum_{j=1}^{M} z_{j} \frac{\partial }
{\partial z_{j}} \biggr\} \Psi_{mn}
\end{displaymath}

\begin{displaymath}
= \biggl\{ - \frac{N^2}{48}(N-\frac{19}{N} + \frac{24}{N^2})+ 
m(\frac{N}{2} - 1 - m) + n(\frac{N}{2} - 1 - n)
\end{displaymath}

\begin{equation}
+ \frac{m-n}{2} \biggr\} \Psi_{mn} - \sum_{\ell = 0}^{\ell_M} (m-n+ 2\ell ) \; \Psi_{m+\ell, n-\ell}
\; \; \; ,
\label{eig4}
\end{equation}

\noindent
where $\ell_M = n$ if $m+n < M$,  $\ell_M = M - m$ if $m + n \geq M$ and, 
in deriving Eq.(\ref{eig4}) we used the identity 
\[
\frac{x + y}{x - y} (x^{m} y^{n} - x^{n} y^{m})
\]

\beq
=  2 \sum_{\ell = 0}^{m-n} x^{m-\ell} y^{n+\ell}
- (x^{m} y^{n} + x^{n} y^{m}) \; \; \; .
\eneq

\noindent
We look for solutions to Eq.(\ref{eig4}) which are linear combinations
of the states $\Psi_{ m+\ell, n-\ell}$:

\beq
\Phi_{mn} = \sum_{\ell=0}^{\ell_M} a^{mn}_l \Psi_{ m +\ell , n-\ell} \;\;\; .
\label{enera}
\label{tras1}
\eneq
\noindent
The coefficients $a_\ell$ are found to be:
 basis
of the states  $ | \Psi_m \rangle$

\begin{equation}
a_{\ell}^{mn} =  \frac{ - ( m - n + 2 \ell ) \;
 }{ 2 \ell ( \ell + m - n + 1/2 )} \sum_{ k=1}^\ell   a^{mn}_{ k - 1}
; \; \; \;   (a_0 = 1)
\label{fmn}
\end{equation}

\noindent
and the corresponding two-spinon energies are given by:

\begin{displaymath}
E_{mn}=-J \frac{\pi^2}{24}(N-\frac{19}{N} + \frac{24}{N^2})  +
\end{displaymath}

\begin{equation}
 \frac{J}{2} (\frac{2\pi}{N})^2 \left[ m(\frac{N}{2} - 1 - m) 
+ n(\frac{N}{2} - 1 - n)
- \frac{m-n}{2}\right] \;\;\; .
\label{pula}
\end{equation}
\noindent

In terms of spinon momenta, the expression of the energy is:

\begin{displaymath}
E_{mn} = - J ( \frac{\pi^2}{24} ) (  N + \frac{5}{N} ) 
\end{displaymath}

\beq
+
[ E (q_m ) + E ( q_n) - \frac{ \pi J}{N} \frac{|q_m - q_n|}{2} ]
 \; \; \;  (q_m\leq q_n) \; \; . 
\label{2spen}
\eneq

\noindent
$E_{mn}$ is the sum of the ground-state contribution, $E_{GS} = - J (
\pi^2 / 24 ) ( N + 5 / N )$, and $E (q_m , q_n)$, which is the two-spinon
energy above the ground state. $E (q_m , q_n)$ is the sum of the energies
of two isolated spinons plus a negative interaction contribution that
becomes negligibly small in the thermodynamic limit.

Such a simple solution for the two-spinon problem 
is possible because the matrix to which Eq.(\ref{eig4}) corresponds is 
lower triangular, i.e. takes the form

\begin{equation}
{\rm Matrix} = \left[ \begin{array}{lllll}
    E_{0}  &   0    &   0    &  0     & \ldots \\
    v_{10} & E_{1}  &   0    &  0     &        \\
    v_{20} & v_{21} & E_{2}  &  0     &        \\
    v_{30} & v_{31} & v_{32} & E_{3}  & \ldots \\
    \vdots &        &        & \vdots &        \end{array} \right]
\; \; \; .
\label{hfmn}
\end{equation}
\noindent
where:

\begin{displaymath}
E_j = \lambda_{ m + j , n - j} \; \;\; {\rm and}
\;\;\; v_{pq} = - ( m - n + 2 p + 2 q)
\end{displaymath}
\noindent
and:
\beq
\lambda_{mn} =   m(\frac{N}{2} - 1 - m) + n(\frac{N}{2} - 1 - n)
- \frac{m-n}{2} \;\;\; .
\label{lambdas}
\eneq
\noindent
The eigenvalues of such a matrix are its diagonal elements, and the
corresponding eigenvectors are generated by recursion.  

The transformation in Eq.(\ref{tras1}) can be inverted and it takes the 
form:

\beq
\Psi_{mn} = \sum_{\ell=0}^{\ell_M} b_\ell^{mn} \Phi_{m+\ell, n-\ell} \;\;\; , 
\label{invtra}
\eneq
\noindent
The coefficients $b_{\ell}^{mn}$ can be expressed in a closed-form formula in
terms of the coefficients of the two-spinon wavefunctions. We provide
their expression in Section VI.

\subsection{The Norm}

The squared norm of the state $\Phi_{mn}$ is defined as:

\beq
\langle \Phi_{mn} | \Phi_{mn} \rangle  = \sum_{ z_1 , \ldots , z_M} | \Phi_{mn} ( z_1 , \ldots , z_M ) |^2 \;\; .
\label{normone1}
\eneq
\noindent
As for the one-spinon wavefunction, we calculate the norm of the two-spinon
states by means of mathematical induction. The details of our calculation
are discussed in Appendix E. The basic induction relations are given
by:

\begin{displaymath}
\frac{\langle \Phi_{mn} | \Phi_{mn} \rangle}{\langle \Phi_{{m},{n-1}} 
| \Phi_{{m},{n-1}} \rangle} 
\end{displaymath}
\beq
=
\frac{ ( n - \frac{1}{2} ) ( M - n + \frac{3}{2} ) ( m - n+ 1)^2}{
n ( M - n + 1 ) ( m - n+\frac{3}{2} ) ( m - n + \frac{1}{2} ) } 
\label{recur1}
\eneq

\begin{displaymath}
\frac{\langle \Phi_{mn} | \Phi_{mn} \rangle}{\langle \Phi_{{m-1},{n}} 
| \Phi_{{m-1},{n}} \rangle} 
\end{displaymath}
\beq 
=
\frac{ ( M - m + 1 ) ( m - n + \frac{1}{2} ) ( m - n - \frac{1}{2} ) m}{
( M -m + \frac{1}{2} ) ( m - n )^2 ( m + \frac{1}{2} ) }  \;\; .
\label{recur2}
\eneq
\noindent
From Eq.(\ref{recur1},\ref{recur2}) one gets the formula for the squared norm:

\[
\langle \Phi_{mn} | \Phi_{mn} \rangle = C_M \frac{M+\frac{1}{2}}{\pi}
\frac{ \Gamma [ m - n + \frac{1}{2} ] \Gamma [ m - n + \frac{3}{
2} ] }{ \Gamma^2 [ m - n + 1 ] }  
\]

\beq
\times \frac{ \Gamma [ m + 1 ] \Gamma [ M - m +
\frac{1}{2} ] }{ \Gamma [ m + \frac{3}{2} ] \Gamma [ M - m +1 ] }
\frac{ \Gamma [ n + \frac{1}{ 2} ] \Gamma [ M - n + 1 ] }{ \Gamma [ 
n + 1 ] \Gamma [ M - n + \frac{3}{2} ] } \; .
\label{bigas}
\eneq
\noindent
Eq.(\ref{bigas}) basically agrees with the result quoted in \cite{kato},
although we derived it by making direct use of the operator ${\cal H}_{HS}$ 
(See Appendix E).

\subsection{Spin Current}

The $\Psi_{mn}$ are eigenstates of $\Lambda^z$. Indeed, a manipulation
similar to the one-spinon case yields:

\begin{equation}
\Lambda^z | \Psi_{mn} \rangle =
\biggl\{ \frac{N-2}{2} - m - n \biggr\} |\Psi_{mn} \rangle \;\;\;,
\label{l2sp}
\end{equation}
\noindent
with the eigenvalue given by the sum of the two spinon velocities.
We will just skip the proof of Eq.(\ref{l2sp}) which works
exactly like the proof of Eq.(\ref{l1sp}).

\subsection{Spinon Statistics}
Spinons are semions, i.e. particles obeying 1/2 fractional statistics.
Since the 2-spinon wavefunction $\Psi_{\alpha \beta}$ has the property

\[
\Psi_{\alpha \beta}^* (z_1 , ... , z_{N/2-1}) = (z_\alpha z_\beta)^{1 - N/2}
\Psi_{\alpha \beta} (z_1 , ... , z_{N/2-1}) \; \; \; ,
\]

\noindent
the Berry phase vector potential for adiabatic motion of spinon $\alpha$
in the presence of $\beta$ is

\begin{displaymath}
\frac{1}{2} \biggl[
\frac{  \langle \psi_{\alpha
\beta } | z_\alpha \frac{\partial} {\partial z_\alpha} \psi_{\alpha
\beta } \rangle +
\langle  z_\alpha \frac{\partial} {\partial z_\alpha} \psi_{\alpha
\beta } | \psi_{\alpha \beta} \rangle }{
  \langle  \psi_{\alpha \beta} | \psi_{\alpha \beta} \rangle }
\biggr]
\end{displaymath}

\begin{equation}
= \frac{1}{2} (1 - \frac{N}{2}) \; \; \; .
\end{equation}

\noindent
The phase to ``exchange'' the spinons by moving $\alpha$ all the way
around the loop is thus

\begin{equation}
\Delta \phi = \oint \frac{1}{2} (1 - \frac{N}{2}) \frac{dz_\alpha}{z_\alpha}
= \pm \frac{\pi}{2} i \; \; \; \pmod{2\pi} \; \; \; .
\end{equation}

\noindent
This number is $0$ or $\pi$ for bosons or fermions.
 The number of states available to
$\ell$ $\downarrow$ spinons, determined by counting the number
of distinct symmetric polynomials of the form

\begin{displaymath}
\Phi_{z_{A_1} , ... , z_{A_\ell}} (z_1 , ... , z_{(N-\ell)/2})
\end{displaymath}

\begin{equation}
= \prod_j^{(N - \ell) / 2} (z_j - z_{A_1}) \times ... \times
(z_j - z_{A_\ell}) \; \; \; ,
\end{equation}

\noindent
is

\begin{equation}
{\cal N}_\ell^{\rm semi} =
( \! \! \begin{array}{c} N/2 + \ell/2 \\ \ell \end{array} \! \! )
\; \; \; .
\end{equation}

\noindent
This is just halfway between the numbers

\begin{equation}
{\cal N}_\ell^{\rm fermi} =
( \! \! \begin{array}{c} N/2 \\ \ell \end{array} \! \! )
\; \; \; \; \; \; \; \; \; \; \; \; \; \; \; \;
{\cal N}_\ell^{\rm bose} =
( \! \! \begin{array}{c} N/2 + \ell \\ \ell \end{array} \! \! )
\; \; \; ,
\end{equation}

\noindent
likewise calculated assuming that the number of states available for
one particle is $N/2$.

\section{Scattering resonance}

In this Section we provide one of the key results of our study: the analysis
of the interaction between two spinons. First, we properly define
the real space representation for the two-spinon relative wavefunction. Then,
we study the behavior of the corresponding amplitude as a function of the
spinon separation. 
Here we construct the real space wavefunction for a spinon pair 
and we show that our results provide a clear evidence for spinons
being interacting particles. 

The real space representation for the two-spinon wavefunctions corresponding
to the energy eigenstate $ | \Phi_{mn} \rangle$,  $z_\alpha^m z_\beta^n p_{mn} 
( z_\alpha / z_\beta)$, is defined
by the decomposition of  the state of two localized
spinons at $z_\alpha$ and $z_\beta$, $|\Psi_{\alpha \beta} \rangle$, in 
the basis  of the fully polarized two-spinon eigenstates:

\beq
\Psi_{\alpha \beta }  = 
\sum_{ m = 0}^M \sum_{n=0}^m (-1)^{m+n} z_\alpha^m z_\beta^n p_{ mn}
( \frac{ z_\alpha}{ z_\beta} ) \Phi_{mn}
\; \; \; .
\label{twospinon}
\eneq

\noindent
$|\Phi_{mn} \rangle$ is an eigenstate of ${\cal H}_{HS}$ with eigenvalue
$E_{mn}$. This implies

\beq
\langle \Phi_{mn} | {\cal H}_{HS} | \Psi_{ \alpha \beta} \rangle = 
E_{mn} \langle \Phi_{mn} | \Psi_{ \alpha \beta} \rangle \; \; \; .
\label{dif1}
\eneq
\noindent
From Eq.(\ref{eig3}) we see that $\langle \Phi_{mn} | {\cal H}_{HS}
| \Psi_{ \alpha \beta } \rangle$ can be written as a differential operator
acting on $\langle \Phi_{ mn} | \Psi_{ \alpha \beta} \rangle$. $\Psi_{\alpha
\beta}$ is perfectly defined for any $z_\alpha$, $z_\beta$ on the unit 
circle, so, the differential operator acts on the analytic extension of
 $\langle \Phi_{ mn} | \Psi_{ \alpha \beta} \rangle$ as

\begin{displaymath}
\langle \Phi_{mn} | {\cal H}_{HS} | \Psi_{ \alpha \beta}
\rangle = E_{GS} \langle \Phi_{mn} |  \Psi_{ \alpha \beta}
\rangle +
\end{displaymath} 

\begin{displaymath}
 \frac{J}{2} ( \frac{2 \pi}{N} )^2
\biggl\{ ( M - z_\alpha \frac{ \partial}{ \partial z_\alpha }) z_\alpha
\frac{ \partial}{ \partial z_\alpha} + ( M - z_\beta \frac{ \partial}{
\partial z_\beta} ) z_\beta \frac{ \partial}{ \partial z_\beta} 
\end{displaymath}

\beq
- \frac{1}{2}  \frac{ z_\alpha + z_\beta}{ z_\alpha - z_\beta} ( z_\alpha
\frac{ \partial}{ \partial z_\alpha} - z_\beta \frac{ \partial}{ \partial 
z_\beta}) \biggr\} \langle \Phi_{mn} | \Psi_{ \alpha \beta} \rangle
\; \; \; .
\label{diffe}
\eneq

\noindent
In the differential operator in Eq.(\ref{diffe}) we recognize the sum of the
energies of the two free spinons and a velocity-dependent interaction, which 
diverges at small spinon separations. Eq.(\ref{diffe}) allows for determination
of the exact expression of $p_{mn} ( z_\alpha / z_\beta )$. Indeed, by using 
Eqs.(\ref{twospinon},\ref{dif1},\ref{diffe}), 
we find the following equation for $p_{mn} ( z )$ ($z = z_\beta / z_\alpha$):

\[
z ( 1 - z) \frac{ d^2 p_{mn} }{ d z^2} + \biggl[   \frac{1}{2} - m + n 
\]

\beq
- ( - m + n + \frac{3}{2} ) z \biggr] \frac{ d p_{mn}}{ d z} +
\frac{ m - n }{2} p_{mn} = 0 \;\;\; .
\label{great}
\eneq
\noindent
This equation is a special case of the hypergeometric equation \cite{abram} 
where the parameters $c,b,a$ are given by

\[
c = \frac{1}{2} - m + n \;\;\; 
b = \frac{1}{2} \;\;\; 
a = -m+n \;\;\; .
\]
\noindent
The solution is a hypergeometric series whose regular solution stops at a 
power of $z$ given by $z^{m-n}$, thus becoming the 
``hypergeometric polynomial''

\[
p_{mn} ( z ) = 
\frac{ \Gamma [ m - n + 1]}{ \Gamma [ \frac{1}{2} ] \Gamma [ m - n + \frac{1}{2
} ] }
\]

\beq
\times
\sum_{ k = 0}^{ m-n} \frac{ \Gamma [ k + \frac{1}{2} ] \Gamma [
m - n - k + \frac{1}{2} ] }{ \Gamma [ k + 1 ] \Gamma [ m - n - k + 1] }
z^k
\label{bigboy2}
\eneq

\noindent
The value of the
spinon wavefunction at zero separation between spinons can be computed
by means of general identities among hypergeometric series \cite{abram}. 
It is given by:
\beq
p_{mn} ( 1 ) = \Gamma [1/2] \Gamma [m - n +1] / \Gamma [m - n + 1/2] \; .
\label{pmn1}
\eneq

According to Eq.(\ref{bigboy2}), $|p_{mn} (z)|^2$ is the 
density of probability for two spinons as a function of the
distance between them.

\begin{figure}
\includegraphics*[width=0.87\linewidth]{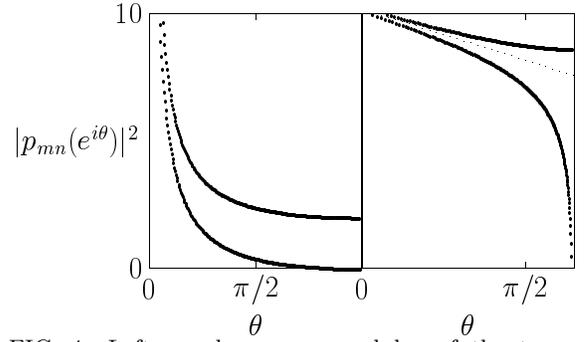}
\caption{Left panel: square modulus of the two-spinon wavefunction as a 
function of the separation between the two spinons for $N=1000$. Right panel:
the same plot on a $\log$-$\log$ scale. The dashed line is a guide to the eye.
It is a plot of $1/x$ with an appropriate offset ($z=\exp (i \theta)$).}
\label{fig5}
\end{figure}

The interpretation of our results is straightforward. Spinons
do actually behave like real particles. Indeed, we have been able to 
determine a differential equation for $p_{mn}$, which for the two-spinon 
wavefunction is the same as the Schr\"odinger equation for a pair of 
ordinary particles. The interaction between spinons is clearly shown
in Fig.(\ref{fig5}), where we  plot $|p_{mn} ( z )|^2$. At large 
separations,  the 
probability density oscillates and averages to 1, independently 
on the distance between the spinons. This is a typical feature of 
noninteracting particles.  Indeed, at large separations spinons are in 
fact noninteracting. However, at small
separations,  $|p_{mn}|^2$ shows a resonant enhancement, 
which corresponds to a huge increase of the probability of configurations 
with the two spinons
on top one of each other. The enhancement is highest at relative spinon 
momentum of $\pi$. Such features are clear evidence of the  
interaction among spinons, which can be characterized as follows:

\begin{enumerate}

\item It is attractive: it favors configurations with spinons on top
one of each other;

\item It is short ranged: spinons are free at large distances.

\end{enumerate}
\noindent 
As $N$ gets larger, the resonant enhancement peaks up \cite{nos}. Hence, 
the resonant enhancement safely survives the thermodynamic limit, 
even though, in this limit, the energy for the two-spinon 
solution is  the
sum of the energies of the two isolated spinons. However, 
the attraction is not strong enough 
to create a two spinon bound state, even in the thermodynamic limit.
This corresponds to the absence of a low-energy stable spin-one excitation, 
which is a typical feature of 1D spin-1/2 antiferromagnets.
 
The attractive force may also be inferred from the energy eigenvalue 
if we rewrite it as

\begin{displaymath}
{\cal H}_{HS} | \Phi_{mn} \rangle 
\end{displaymath}

\[
=
 \biggl\{ - J (\frac{\pi^2}{24}) ( N +
\frac{5}{N} ) + E_{q_{m}} + E_{q_{n}} + V_{q_{m} - q_{n}} \biggr\}
| \Phi_{mn} \rangle =
\]

\beq
\biggl\{ - J (\frac{\pi^2}{24}) ( N +
\frac{5}{N} ) + E ( q_{m} , q_{n} )  \biggr\} | \Phi_{mn} \rangle \;\;\; ,
\label{eig5}
\end{equation}

\noindent
where

\begin{equation}
V_{q} = - J \frac{\pi}{N} \mid \! q \! \mid
\; \; \; .
\end{equation}

\noindent
Note that this potential vanishes as $N \rightarrow \infty$, as
expected for particles that interact only when they are close together.
 However, as we already pointed out, the vanishing of the interaction 
potential in the thermodynamic limit does not mean that no residual effects 
survive such a limit. The resonant enhancement when the two spinons are on the
same site does survive the thermodynamic limit and it is the main reason of
the instability of the spin-1 spin-wave, as we discuss at length in the next
section.

Before concluding this Section, we provide  the expression of the 
coefficients $b_\ell^{mn}$ in Eq.(\ref{invtra}). From Eq.(\ref{bigboy2})
it is straightforward to prove that:

\begin{displaymath}
b_\ell^{mn} = \frac{ \Gamma [ m - n + 2 \ell + 1 ] }{ \Gamma [ \frac{1}{2} ] 
\Gamma [ m - n + 2 \ell + \frac{1}{2} ] }
\end{displaymath}

\beq
\times \frac{ \Gamma [ \ell + \frac{1}{2} ] }{ \Gamma [ \ell + 1 ] } 
\frac{ \Gamma [ m - n + \ell  + \frac{1}{2} ] }{ \Gamma [ m - n + \ell + 1 ]}
\;\;\; .
\label{thebs}
\eneq

\section{Spin Susceptibility}

In this Section, we work out the dynamical spin susceptibility 
for the HSM. We show that the DSS  depends only on the 
$p_{mn}$'s calculated at $z=1$, which allows us to obtain for any
finite $N$ a simple closed-form expression for the DSS and to  
relate it to the spinon interaction. 
By carefully taking the thermodynamic limit of our result, we obtain 
Haldane-Zirnbauer formula for the DSS in the thermodynamic limit \cite{hazi}. 
Haldane-Zirnbauer formula shows that there is no low-energy spin-1 pole 
in the DSS, but
the function takes a sharp square-root singularity at the two-spinon
threshold on top of a branch cut, corresponding
to the lack of integrity of the spin-one excitation. Our analysis  
definitely proves  that the square root sharp edge on top of the broad 
spectrum is nothing but the interaction between spinons. The resonant 
enhancement is the square root singularity in the spin-susceptibility. 
This result  is of the outermost importance, 
since it represents a way to experimentally test interaction among
 spinons in 1D. We will come back to such a point in the concluding remarks.

Let us begin with the calculation of the spin susceptibility for a finite
lattice. The DSS is  the dynamical propagator for a spin-1 spin flip.  A 
spin flip with momentum $q$ is created by acting on $|\Psi_{ GS} \rangle$ with
$S_q^-$, defined as:

\beq
S_q^- = \sum_\alpha (z_\alpha^* )^k (S_\alpha^x - i S_\alpha^y )
\; \; \; \; \; \; \;
(q = 2 \pi k / N ) \; \; \; .
\label{floppi}
\eneq
\noindent
A peculiar property of the HSM is that a spin flip at $z_\alpha$ is the
same as a spinon pair at the same site \cite{hazi}. Therefore, we can
fully decompose $S_q^- \Psi_{ GS}$ in the basis of the two-spinon 
eigenstates:

\[
S_q^- \Psi_{GS} = \sum_\alpha (z_\alpha^* )^k \Psi_{\alpha \alpha}
\]

\beq
= N \sum_{m = 0}^M \sum_{n = 0}^m (-1)^{m+n}  p_{mn}(1)
\delta(m+n-k) \; \Phi_{mn} \; \; \; .
\label{flippi}
\eneq

The susceptibility is given by

\[
\chi_q(\omega) = \sum_X
\frac{| \langle X | S_q^- | \Psi_{GS} \rangle |^2}
{\langle X | X \rangle \; \langle \Psi_{GS} | \Psi_{GS} \rangle }
\]

\beq
\times
\frac{2 (E_X - E_{GS})}{(\omega + i \eta)^2 - (E_X - E_{GS} )^2}
\; \; \; ,
\label{gennaro}
\eneq

\noindent
( $| X \rangle$ is an exact eigenstate of ${\cal H}_{HS}$ with energy
$E_{X}$). Then, from Eqs.(\ref{flippi},\ref{gennaro}) we  have that $\chi_q (
\omega)$ takes a nonzero contribution only if $| X \rangle = | \Phi_{mn}
\rangle$. Then, Eq.(\ref{gennaro}) becomes:
 
\[
\chi_q (\omega) = N^2 \sum_{m = 0}^M \sum_{n = 0}^m
\frac{\langle \Phi_{mn} | \Phi_{mn} \rangle }
{ \langle \Psi_{GS} | \Psi_{GS} \rangle } p_{mn}^2 (1)
\]

\beq
\times \delta (m + n - k ) \;
\frac{2 (E_{mn} - E_{GS} )}{(\omega + i \eta)^2
- (E_{mn} - E_{GS} )^2 } \; \; \; .
\label{entirely}
\eneq

\noindent
Eq.(\ref{entirely}) is another relevant result of our work. It shows that
only the $p_{mn} ( z)$'s at $z=1$ determine $\chi_q ( \omega  )$. Therefore,
the spin susceptibility is completely determined by spinon interaction.

Let us analyze the thermodynamic limit of eq.(\ref{entirely}). In the
thermodynamic limit  the gamma functions can be approximated by using 
Stirling's formula:

\beq
\Gamma [ z ] \approx \sqrt{\pi} ( z - 1 )^{ ( z - \frac{1}{2} )} e^{ - ( z- 1
)} \;\;\; , 
\label{stirl}
\eneq

\noindent
From Eqs.(\ref{pmn1},\ref{stirl}) we get, in the thermodynamic limit:

\begin{displaymath}
N^2\frac{\langle \Phi_{mn} | \Phi_{mn} \rangle }
{ \langle \Psi_{GS} | \Psi_{GS} \rangle } p_{mn}^2 (1)=
\end{displaymath}

\beq
\frac{\pi N(m-n+\frac{1}{2})}{\sqrt{n(M-n)(m+\frac{1}{2})(M-m-\frac{1}{2})}}
\;\;\; .
\label{order1}
\eneq

\noindent
Since the joint two-spinon density of states is flat,  
the sums over $m$ and $n$ become integrals over the (halved) one-spinon
Brillouin zone:

\beq
\sum_m \rightarrow - M \int_{-\frac{ \pi}{2}}^\frac{ \pi}{ 2} \frac{ d q}{ \pi}
\label{flat}
\eneq

\noindent
From Eq.(\ref{stirl},\ref{order1},\ref{flat}), we see that 
Eq.(\ref{entirely}) turns, in
the thermodynamic limit, into the Haldane-Zirnbauer formula for the DSS 
\cite{hazi}

\[
\chi_q ( \omega ) = \frac{J}{ 2 } \int_{ - \frac{ \pi}{2}
}^\frac{\pi}{
2} d q_1  \int_{ - \frac{  \pi}{2} }^{ q_1} 
d q_2\frac{ | q_1 - q_2 |\delta ( q_1 + q_2 - q  )
 }{ \sqrt{ E ( q_1 ) E ( q_2 
) } } 
\]

\beq
\times \frac{2 E(q_1 , q_2 ) }
{(\omega + i \eta)^2 - E^2 (q_1 , q_2 )}
\; \; \; ,
\label{ter}
\eneq

\noindent
where $E ( q )$ and $E ( q_1 , q_2 )$ are the one-spinon and the
two-spinon energies, respectively.  Integration pver $q_1$, $q_2$ in 
Eq.(\ref{ter}) provides:

\[
\chi_q ( \omega ) =  \frac{J}{4}
\]

\beq
\times \frac{\Theta [ \omega_2 (q ) - \omega ] \;
\Theta [ \omega - \omega_{-1} ( q ) ] \; 
\Theta [ \omega - \omega_{ +1} ( q )]}
{\sqrt{ \omega - \omega_{ -1} (q )}
\sqrt{ \omega - \omega_{ +1} ( q ) }} \; \; \; ,
\label{the2}
\eneq

\noindent
$\omega_{-1} ( q )$ and $\omega_{+1} ( q )$ are the threshold energies
for a spinon pair with momentum $q$,  according to whether 
$0 \leq q \leq \pi$ or $\pi \leq q \leq 2 \pi$, respectively. They are
given by $ \omega_{-1} ( q ) = (J/2) q ( \pi - q )$, $\omega_{+1} ( q ) =
(J/2) ( 2 \pi - q )( q - \pi )$. $ \omega_2 ( q ) = (J/2) q ( 2 \pi - q)$
is the upper threshold for the spin-1 excitation. From Eq.(\ref{the2})
we see that the resonant enhancement, given by $p_{mn}^2 ( 1 )$, has turned
into a square-root singularity  in $\chi_q ( \omega )$ vs. 
$\omega$ at fixed $q$, with the branch cut originating either at 
$\omega_{-1} ( q)$ or at $\omega_{+1} ( q )$, depending on the value of $q$.
Because the two-spinon joint density of states is uniform, the main 
conclusion we trace out from our calculation is that the branch cut in 
$\chi_q ( \omega )$, i.e., the broadness of the spectral density of states, 
is the spinon interaction. 

A measurement of $\chi_q ( \omega )$ in one-dimensional
spin-1/2 antiferromagnets can be performed my means of neutron
scattering experiments \cite{tennant}. The result of the measurements 
\cite{tennant} does, 
in fact, show a sharp threshold followed by a broad spectrum, 
in good agreement with predictions  of Eq.(\ref{the2}). In light of our 
present discussion,  we conclude that what is actually seen in such an 
experiment is  a direct consequence of the spinon interaction in 
1-dimensional antiferromagnets.
Hence, the experiments provide evidence that spinons do interact and that 
 the spinon interaction is what determines the peculiar low-energy 
physics of spin-1/2 antiferromagnetic  chains.

From Eq.(\ref{flippi}) we also derive  the formula for the spin-spin 
correlation function, $\chi ( z_\alpha )$, in terms of the two-spinon 
wavefunction at $z=1$:

\[
\chi ( z_\alpha ) = \frac{ \langle \Psi_{ GS } | S_1^+ S_{ z_\alpha}^- | 
\Psi_{ GS} \rangle }{ \langle \Psi_{ GS} | \Psi_{ GS } \rangle} 
\]

\beq
=
\sum_{ m = 0}^M \sum_{ n = 0 }^m ( z_\alpha )^{ m + n } 
\frac{ \langle \Phi_{ mn} | \Phi_{ mn} \rangle }{ 
\langle \Psi_{ GS } | \Psi_{GS} \rangle } 
p_{mn}^2 ( 1 ) \;\;\; .
\label{static}
\eneq

Eq.(\ref{static}) is the formula we have plotted in Fig.(\ref{fig2}).

\section{Conclusions}

In this paper we developed a simple approach to the study of spinon
excitations of the Haldane Shastry model, based on the formalism of the
analytic variables. Within out approach we picture spinons as local 
defects in the disordered sea. Our formalism allows for a 
consistent real-space representation of the wavefunction for two spinons.
We construct the Schr\"odinger equation, whose solution is the two-spinon
wavefunction, which  shows that spinons behave as real quantum-mechanical 
particles. By means of a  careful study of the real-space two-spinon 
wavefunction, we reveal the main result: the existence of spinon interaction 
and its survival in the thermodynamic limit. 
Spinon interaction is a short-range attraction, which generates a resonant 
enhancement of the probability for two spinons to be at the same site. 
Such an interaction determines  the low-energy physics of
1D interacting antiferromagnets. Since the low-energy joint density of
states is uniform, the broadness in the spectral density, 
 is exclusively caused by the resonant enhancement, as  
we show from  the finite-$N$ expression for the spin susceptibility 
(Section VII).  In the thermodynamic limit the resonant enhancement 
develops a square root singularity followed by a branch cut, which is the 
broadness in the spectral density of
states. The branch cut reflects the absolute instability of the spin wave
towards decay into a spinon pair. Then, we show that, 
even though in the thermodynamic limit the interaction  is irrelevant, 
its main effect, the resonant enhancement, peaks up. 

In conclusion, we analyzed spinon interaction in an exact solution of
the Haldane-Shastry model and  its consequences for the low-energy physics of
1-D spin-1/2 antiferromagnets.

\section*{Acknowledgments}

This work was supported primarily by the National Science
Foundation under grant No. DMR-9813899. Additional support was provided by
the U.S. Department of Energy under contract No. DE-AC03-76SF00515 and by
the Bing Foundation.

\newpage

\appendix

\section{Fourier Sums}

In this Appendix we will prove some of the formulas we used throughout 
the paper.
\noindent
Since the lattice sites $z_\alpha$ are roots of unity we have

\begin{equation}
\prod_\alpha^N (z - z_\alpha ) = z^N - 1 \; \; \; .
\end{equation}

\noindent
Then for $0 \leq m \leq N$ we have

\begin{figure}
\includegraphics*[width=0.87\linewidth]{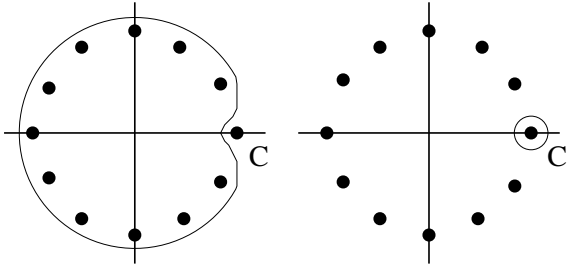}
\caption{Contours used in Eqs(\ref{first}) and (\ref{second}).}
\label{fc}
\end{figure}

\begin{displaymath}
\sum_{\alpha = 1}^{N-1} \frac{z_\alpha^m}{z_\alpha - 1}
\end{displaymath}

\begin{displaymath}
= \frac{N}{2\pi i} \oint_C \frac{z^{m-1} \; dz}{(z - 1)(z^N - 1)}
= - \frac{N}{2\pi i} \oint_{C'} \frac{z^{m-1} \; dz}{(z - 1)(z^N - 1)}
\end{displaymath}

\begin{displaymath}
= - \frac{N}{2\pi i} \oint \biggl\{ \frac{1 +
( \! \! \begin{array}{c} m-1 \\ 1 \end{array} \! \! ) \; x +
( \! \! \begin{array}{c} m-1 \\ 2 \end{array} \! \! ) \; x^2 + ... }
{ ( \! \! \begin{array}{c} N \\ 1 \end{array} \! \! ) \; +
( \! \! \begin{array}{c} N \\ 2 \end{array} \! \! ) \; x +
( \! \! \begin{array}{c} N \\ 3 \end{array} \! \! )\; x^2 +
... } \biggr\} \; \frac{dx}{x^2}
\end{displaymath}

\begin{equation}
= \frac{N+1}{2} - m
\; \; \; ,
\label{first}
\end{equation}

\noindent
and

\begin{displaymath}
\sum_{\alpha = 1}^{N-1} \frac{z_\alpha^m}{|z_\alpha - 1|^2}
= - \sum_{\alpha = 1}^{N-1} \frac{z_\alpha^{m+1}}{(z_\alpha - 1)^2}
\end{displaymath}

\begin{displaymath}
= - \frac{N}{2\pi i} \oint_C \frac{z^m \; dz}{(z - 1)^2(z^N - 1)}
= \frac{N}{2\pi i} \oint_{C'} \frac{z^m \; dz}{(z - 1)^2(z^N - 1)}
\end{displaymath}

\begin{displaymath}
= \frac{1}{2\pi i} \oint \biggl\{ \frac{1 +
( \! \! \begin{array}{c} m-1 \\ 1 \end{array} \! \! ) \; x +
( \! \! \begin{array}{c} m-1 \\ 2 \end{array} \! \! ) \; x^2 + ... }
{ ( \! \! \begin{array}{c} N \\ 1 \end{array} \! \! ) \; +
( \! \! \begin{array}{c} N \\ 2 \end{array} \! \! ) \; x +
( \! \! \begin{array}{c} N \\ 3 \end{array} \! \! ) \; x^2 +
... } \biggr\} \; \frac{dx}{x^3}
\end{displaymath}

\begin{equation}
= \frac{N^2-1}{12} - \frac{m(N-1)}{2} + \frac{m(m-1)}{2}
\; \; \; .
\label{second}
\end{equation}

\section{Calculations of the coefficients $A_l$.}

In this Appendix we work out the  coefficients $A_l$ that appear in the 
eigenvalue equations for the eigenfunctions of the HSM are defined as:

\begin{equation}
A_l = - \sum_{\alpha =1}^{N-1} \frac{z_\alpha^2}{ ( z_\alpha - 1)^{ 2-l}}\;\;\;
,
\end{equation}
\noindent
They can be computed by using the equations from Appendix A. In
particular we have:

\begin{displaymath}
A_0 =- \sum_{\alpha =1}^{ N-1} \frac{ z_\alpha^2}{
( z_\alpha-1)^2} \;\;\; ,
\end{displaymath}

\begin{displaymath}
\sum_{\alpha =1}^{ N -1} \frac{ z_\alpha}{ | z_\alpha -1 |^2} =
\frac{ ( N-1)( N-5)}{ 12} \;\;\; ,
\end{displaymath}

\begin{displaymath}
A_1 = - \sum_{\alpha = 1}^{N-1} \frac{ z_\alpha^2}{ z_\alpha -1}
= - \frac{N-3}{2} \;\;\; ,
\end{displaymath}

\begin{displaymath}
A_2 = - \sum_{\alpha =1}^{ N -1} z_\alpha^2 = 1 \;\;\; ,
\end{displaymath}

\begin{equation}
A_l = - \sum_{ \alpha =1}^{ N -1} z_\alpha^2 ( z_\alpha -1)^{ l-2}
=0 \;\; \; \; ( l > 2) \;\; .
\label{coef1}
\end{equation}

\section{The spin current}

In this Appendix we provide the physical interpretation of the operator
$\vec{\Lambda}$, as the spin-current operator. In order to do so,
 we first construct the continuous interpolation 
of the lattice spin field, given
by the spin density operator $\vec{\rho} ( z )$. Then, we define
a current density on the unit circle, $\vec{j} ( z )$. We prove that
$\vec{\rho}$ and $\vec{j}$ obey an equation which, once restricted to
the lattice, takes the form of the continuity equation for the spin
density. The operator $\vec{\Lambda}$ comes out to be the global
operator whose density is given by $\vec{j} ( z )$.

The first step of such a construction is defining the field interpolating 
the spin operators into the interstices by means of the formula

\begin{equation}
\vec{\sigma}(z) = \biggl[ \frac{z^{N/2} - z^{-N/2}}{2N}\biggr]
\sum_\beta^N \biggl( \frac{z + z_\beta}{z - z_\beta}
\biggr) \vec{S}_\beta \; \; \; .
\label{sigma}
\end{equation}

\noindent
Then we can associate to $\vec{\sigma} ( z )$
a ``$\sigma$ model-like'' Hamiltonian given by:

\begin{displaymath}
\frac{1}{2\pi i} \oint
[ z \frac{d \vec{\sigma}}{d z} ] \cdot [ z \frac{d \vec{\sigma}} {d z} ]
\frac{dz}{z} = - \frac{2}{N} \sum_{\alpha \neq \beta}^N
\frac{\vec{S}_\alpha \cdot \vec{S}_\beta}{| z_\alpha - z_\beta |^2}
\end{displaymath}

\begin{equation}
+ \frac{3}{8}(N - 1) + \frac{S^2}{8}
\label{hsm}
\end{equation}

\noindent
where the integral is performed over the unit circle. Eq.(\ref{hsm}) gives 
the HS Hamiltonian plus an irrelevant constant and an operator which commutes 
with it.

 We also have spin density and spin current density operators

\begin{displaymath}
\vec{\rho}(z) = - i \vec{\sigma}(z) \times \vec{\sigma}(z)
\end{displaymath}

\begin{equation}
\vec{j} (z) =
\frac{1}{2i} \biggl\{ \vec{\sigma} \times [ z \frac{d \vec{\sigma}}{d z}]
- [ z \frac{d \vec{\sigma}}{d z} ] \times \vec{\sigma} \biggr\}
\; \; \; ,
\end{equation}

\noindent
That $\rho ( z )$ is an appropriate
 definition of the spin density may be seen
by taking the limit $z \rightarrow z_\alpha$, being $z_\alpha$ a site
on the lattice. One gets

\begin{equation}
\lim_{ z \rightarrow z_\alpha } \vec{ \rho } ( z ) = \vec{S}_\alpha
\;\;\; .
\end{equation}

\noindent
That $\vec{j} ( z ) $ is a proper spin current can be inferred from the
continuity equation

\begin{equation}
\lim_{z \rightarrow z_\alpha} \; \biggl\{  z \frac{d \vec{j}}
{d z} + [ \sum_{\alpha \neq \beta}^N \frac{\vec{S}_\alpha \cdot
\vec{S}_\beta}{| z_\alpha - z_\beta |^2 }
, \vec{\rho} ] \biggr\} = 0 \;\;\; .
\end{equation}

\noindent
The zero-momentum component of this conserved current density is

\begin{equation}
\frac{1}{2\pi i} \oint \vec{j} \; \frac{dz}{z} = \vec{\Lambda} \;\;\; .
\end{equation}

\noindent
The operator $\vec{\Lambda}$ is then a scaled spin current. Its action on
the state with a fixed number of propagating spinons, 
Eqs.(\ref{lgss},\ref{l1sp},\ref{l2sp}), is
definitely consistent with such an interpretation.

\section{Factorizability of ${\cal H}_{HS}$}

In this Appendix we will prove the factorization formula, Eq.(\ref{factri}). 
In order to do so, we split the proof in two steps. First, we will show 
that the  operator $\vec{D}_\alpha$ annihilate $| \Psi_{ GS} \rangle$, 
then, we will prove the factorization equation. 
\noindent
Let us begin with the first proof.
The operator

\[
\Omega_\alpha = \sum_{\beta \neq \alpha}^N \frac{ z_\alpha}{ z_\alpha -
z_\beta} [ S_\alpha^+ S_\beta^- - ( S^z_\alpha + \frac{1}{ 2} ) ( 
S^z_\beta + \frac{1}{2} ) 
\]

\beq- \frac{ N - 1}{ 2} ( S^z_\alpha + \frac{1}{2} 
) ]
\eneq
\noindent
annihilates $|\Psi_{GS} \rangle$. Indeed, by using the technique we 
developed in Section III, we find

\[
[ \Omega_\alpha  \Psi_{GS} ] ( z_1 , \ldots , z_M )  
\]

\begin{displaymath}
=
 \sum_{\ell = 0}^{N-2} \biggl\{\frac{1}{\ell!}
\sum_{\beta \neq \alpha}^N \frac{z_\alpha z_\beta
(z_\beta - z_\alpha )^\ell} {z_\alpha - z_\beta} \biggr\}
\frac{\partial^\ell}{\partial z_\alpha^\ell} \biggl\{
\frac{\Psi_{GS} (z_\alpha , \dots , z_{N/2} )}{z_\alpha} \biggr\} 
\end{displaymath}

\[
-
 \sum_{\beta \neq \alpha}^N \frac{z_\alpha}
{z_\alpha - z_\beta} \biggl[ - 2 (\frac{1}{2}
+ S_\alpha^z ) (\frac{1}{2} + S_\beta^z ) + \frac{N-1}{2} \biggr] 
\]

\beq
\times
\Psi_{GS} ( z_1 , \ldots , z_M ) = 0 \;\;\; .
\eneq

\noindent
for all $\alpha$.  However since $| \Psi_{GS} \rangle$ is a spin singlet
the irreducible representations of the rotation group present
in this operator must destroy $| \Psi_{GS} \rangle$ separately.  The scalar
component is identically zero.  The vector component is

\begin{equation}
\sum_{\beta \neq \alpha}^N \frac{z_\alpha}{z_\alpha - z_\beta}
[ i (\vec{S}_\alpha \times \vec{S}_\beta ) + \vec{S}_\beta ]
| \Psi_{GS} \rangle = 0 \; \; \; .
\end{equation}

\noindent
Since $| \Psi_{GS} \rangle$ is also its own
time-reverse it must be destroyed by the time-reverse of
the vector operator, i.e.

\[
\sum_{\beta \neq \alpha}^N \frac{z_\alpha^*}
{z_\alpha^* - z_\beta^*} [ i (\vec{S}_\alpha \times \vec{S}_\beta )
+ \vec{S}_\beta ] =
\]

\beq
 - \sum_{\beta \neq \alpha}^N \frac{z_\beta}
{z_\alpha - z_\beta} [ i (\vec{S}_\alpha \times \vec{S}_\beta )
+ \vec{S}_\beta ] \; \; \; .
\eneq

\noindent
The difference of these is the trivial operator $\vec{S}_\alpha \times
\vec{S}$, and their sum is $2 \vec{D}_\alpha$.

We prove, now, the factorizability of ${\cal H}_{HS}$. In order to do so, we 
need the following identities:

\[
\sum_\alpha^N [i (\vec{S} \times \vec{S}_\alpha) + \vec{S} ] \cdot
\vec{D}_\alpha 
\]

\beq
=
\sum_\alpha^N [i \vec{S} \cdot (\vec{S}_\alpha
\times \vec{D}_\alpha ) + \vec{S} \cdot \vec{D}_\alpha ]
= \frac{3}{2} \vec{S} \cdot \vec{\Lambda} \; \; \; ,
\end{equation}

\begin{displaymath}
\sum_\alpha^N [ i ( \vec{S} \times \vec{S}_\alpha ) + \vec{S} ] \cdot
[ i (\vec{S}_\alpha \times \vec{S} ) + \vec{S} ]
\end{displaymath}

\begin{equation}
 = \frac{3}{2} [ N - 1 ] S^2 \; \; \; ,
\end{equation}

\begin{displaymath}
\sum_{\beta \neq \gamma \neq \alpha}^N \frac{
\vec{S}_\beta \cdot \vec{S}_\gamma}{ ( z^*_\alpha - z^*_\gamma )( z_\alpha -
z_\beta ) } 
\end{displaymath}

\begin{equation}
= -\frac{1}{2} S^2 + \frac{3}{8} N + 2\sum_{\alpha \neq \beta}^N
\frac{\vec{S}_\alpha \cdot \vec{S}_\beta}{|z_\alpha - z_\beta|^2}
\; \; \; .
\end{equation}

\begin{displaymath}
i\sum_{\alpha \neq \beta \neq \gamma}^N \frac{\vec{S}_\gamma \cdot
(\vec{S}_\alpha \times \vec{S}_\beta)}{(z^*_\alpha - z^*_\gamma) ( z_\alpha -
z_\beta)}
\end{displaymath}

\begin{displaymath}
= i \sum_{\alpha \neq \beta \neq \gamma}^N \frac{z_\alpha z_\gamma}
{ ( z_\alpha - z_\gamma ) ( z_\alpha - z_\beta ) }
\vec{S}_\alpha \cdot (\vec{S}_\gamma \times
\vec{S}_\beta)
\end{displaymath}

\begin{equation}
= \frac{i}{2} \sum_{\alpha \neq \beta \neq \gamma}^N \biggl(
\frac{z_\alpha + z_\gamma}{z_\alpha - z_\gamma} \biggr)
(\vec{S}_\alpha \times \vec{S}_\gamma) \cdot \vec{S}_\beta
= \vec{\Lambda} \cdot \vec{S}
\; \; \; .
\end{equation}

\noindent
By putting together the identity

\begin{displaymath}
\sum_\alpha^N \sum_{\beta \neq \alpha}^N \sum_{\gamma \neq \alpha}^N \frac{
[i (\vec{S}_\alpha \times \vec{S}_\gamma) + \vec{S}_\gamma ]^\dagger
\cdot [i (\vec{S}_\alpha \times \vec{S}_\beta) + \vec{S}_\beta ]}
{(z_\alpha^* - z_\gamma^*)(z_\alpha - z_\beta )}
\end{displaymath}

\begin{equation}
= \sum_\alpha^N \vec{D}_\alpha^\dagger \cdot \vec{D}_\alpha
+ \frac{3}{2} \vec{S} \cdot \vec{\Lambda} + \frac{3}{8} (N - 1) S^2
\; \; \; 
\end{equation}

\noindent
and the identity

\begin{displaymath}
\sum_{\alpha}^N \sum_{\beta \neq \alpha}^N \sum_{\gamma \neq \alpha}^N
\frac{[ -i(\vec{S}_\alpha \times \vec{S}_\gamma) + \vec{S}_\gamma] \cdot
[i (\vec{S}_\alpha \times \vec{S}_\beta) + \vec{S}_\beta ]}
 {(z^*_\alpha - z^*_\gamma)(z_\alpha - z_\beta)}
\end{displaymath}

\begin{displaymath}
= \frac{3}{2} \sum_\alpha^N \sum_{\beta \neq \alpha}^N \sum_{\gamma \neq
\alpha}^N \frac{1}{(z^*_\alpha - z^*_\gamma)(z_\alpha - z_\beta)}
[ \vec{S}_\gamma \cdot \vec{S}_\beta + i \vec{S}_\gamma \cdot
(\vec{S}_\alpha \times \vec{S}_\beta)]
\end{displaymath}

\[
= \frac{3}{2} \biggl[ 3 \sum_{\alpha \neq \beta} \frac{\vec{S}_\alpha
\cdot \vec{S}_\beta}{|z_\alpha - z_\beta|^2} + \frac{N(N^2 + 5)}{16}
\]

\beq
- \frac{S^2}{2} + \vec{S} \cdot \vec{\Lambda} \biggr]
\; \; \; ,  
\end{equation}
\noindent
the proof is complete.

\bigskip

\noindent
Since $\langle \Phi | \vec{D}_\alpha^\dagger \cdot
\vec{D}_\alpha | \Phi \rangle$ is nonnegative for any wavefunction
$| \Phi \rangle$, this provides an explicit demonstration that
$| \Psi_{GS} \rangle$ is the true ground state.  The annihilation operators
and their equivalence to ${\cal H}_{HS}$ when squared and summed were
originally discovered by Shastry \cite{shastry1}.  They are lattice
versions of the Knizhnik-Zamolodchikov operators known from studies
of the Calogero-Sutherland model, the 1-dimensional Bose gas with
inverse-square repulsions \cite{zam,calogero,sutherland}.

\section{The norm of the states}

In this Appendix we provide the proof of the formula for the norm of the
one-spinon and of the two-spinon eigenstates.

Throughout all this section and the following one, we will make use
of the scalar product between symmetric polynomials $f ( z_1 , \ldots , z_M)$,
defined as

\beq
\langle f | g \rangle=
\sum_{ z_1 , \ldots , z_M} f^*(z_1,
\ldots , z_M ) g ( z_1 , \ldots , z_M ) \;\;\; .
\label{scpro}
\eneq

\noindent
Let us begin with the one-spinon eigenstates. The state for one spinon in
coordinate space, $\Psi_\alpha ( z_1 , \ldots , z_M )$, has been defined in
Eq.(\ref{spinon1}), in the odd-$N$ case ($M=(N-1)/2$) to be:

\beq
\Psi_\alpha (z_1 , \ldots, z_M) = \prod_j^M ( z_\alpha - z_j ) 
\prod_{i<j}^M ( z_i - z_j)^2 \prod_j^M z_j \;\; .
\eneq

\noindent
$\Psi_\alpha$ is of the form $\Phi_\alpha \times \Psi_{GS} $, where 
$\Phi_\alpha = \prod_j^M ( z_\alpha - z_j )$. 

We will prove the formula for the norm of the state $| \Phi_m \rangle$, 
Eq.(\ref{norma2}), by mathematical induction. In order to do so, 
let us define the symmetric operator:

\beq
e_1 ( z_1 , \ldots , z_M ) = z_1 + \ldots + z_M \;\;\; .
\label{symo}
\eneq

\noindent
For any wavefunction of the form $\Phi \times \Psi_{GS}$, where $\Phi$ is 
a symmetric polynomial with degree less than $N-2 M + 2$, we have

\begin{displaymath}
{\cal H}_{HS} \Phi  \Psi_{GS} = E_{GS} \Phi \Psi_{GS} +
\frac{J}{2} (\frac{2\pi}{N})^2 \Psi_{GS} \biggl\{ \frac{1}{2} \biggl[ 
\sum_j z_j^2 \frac{\partial^2}{\partial z_j^2}
\end{displaymath} 

\begin{equation}
+ 4 \sum_{j \neq k}
\frac{z_j^2}{z_j - z_k} \frac{\partial}{\partial z_j} \biggr] - 
\frac{N - 3}{2} 
\sum_j z_j
\frac{\partial}{\partial z_j} \biggr\} \Phi
\; \; \; ,
\end{equation}

\noindent
and thus

\begin{displaymath}
{\cal H}_{HS} e_1 \Phi \Psi_{GS} - e_1 {\cal H}_{HS} \Phi \Psi_{GS} =
\frac{J}{2} (\frac{2\pi}{N})^2 
\end{displaymath}

\begin{equation}
\times \Psi_{GS} \biggl[ \sum_j z_j^2 \frac{\partial}{\partial z_j}
+ \frac{N-3}{2} e_1 \biggr] \Phi \; \; \; .
\label{eqe5}
\end{equation}

\noindent

From the definition of $\Phi_\alpha$ one then obtains:

\beq
\sum_{ j}^M z_j^2 \frac{ \partial}{ \partial z_j} \Phi_\alpha =
[ e_1  + M z_\alpha - z_\alpha^2 \frac{ \partial}{ 
\partial z_\alpha } ]  \Phi_\alpha \;\;\; .
\label{comm2}
\eneq
\noindent
Eq.(\ref{comm2}) implies the following identity for $\Psi_m$:

\[
[  {\cal H} - \frac{ E_{GS}}{ (  J  2 \pi^2 ) /N^2 } ] 
 [ e_1 \Psi_m ] = [ m ( M - m ) + M ] [ e_1 \Psi_m ]  
\]

\beq
 +
 [ M - ( m - 1 ) ] 
\Psi_{ m - 1 } \;\;\; . 
\label{stuf1}
\eneq 
\noindent
(${\cal H}$ is the scaled Hamiltonian: ${\cal H} = {\cal H}_{HS} /
( J 2 \pi^2 / N^2$)).
\noindent
Then, the following identity chain is proved:

\[
(m-1)(M-m+1) \langle \Psi_{ m-1} | e_1 | \Psi_m  \rangle 
\]

\[
=
\langle \Psi_{m-1} | [ {\cal H} - \frac{ E_{GS}}{ (  J  2 \pi^2 ) /N^2 } ] 
 | [e_1  \Psi_m]  \rangle  
\]

\[
=
( M + m ( M - m ) ) \langle \Psi_{ m-1} | e_1 |  \Psi_m  \rangle 
\]
\beq
+
( M - m +1 ) \langle \Psi_{ m -1} | \Psi_{ m-1} \rangle \;\;\; ,
\label{stuf2}
\eneq
\noindent
thus, Eq.(\ref{stuf2}) implies the identity:

\[
\langle \Psi_{m-1} | e_1 |  \Psi_m  \rangle = 
\]

\beq
- \frac{ M - m +1}{ 2 ( M - m + \frac{1}{2} )} \langle 
\Psi_{ m-1} | \Psi_{ m-1} \rangle \;\;\; .
\label{res1}
\eneq
\noindent
In order to determine a suitable induction relation, let us introduce
the operator

\[
e_M ( z_1 , \ldots , z_M ) = z_1 \cdot \ldots \cdot z_M \;\;\; .
\]
\noindent
Clearly

\beq
e_M ( z_1 , \ldots , z_M ) e_M ( \frac{1}{ z_1} , \ldots , \frac{1}{ z_M }) 
= 1 \;\;\; .
\label{fund2}
\eneq
\noindent
Since all the $\Psi_m$'s are products of the ground state factor, 
$\Psi_{GS}$, times a symmetric polynomial of degree less than 2 in
each variable, we have

\[
\langle  \Psi_{ m-1} | e_1 | \Psi_m  \rangle = \langle
[ e^*_M \Psi_m ] | e_1 | [ e^*_M  \Psi_{ m - 1 } ] \rangle
\]

\beq
=
\langle \Psi_{ M - m } |  e_1 | \Psi_{ M - m + 1 }  \rangle \;\;\; .
\label{stuf3}
\eneq

\noindent
At this point, we use again Eq.(\ref{res1}) in order to write Eq.(\ref{stuf3})
as

\[
\langle \Psi_{ M - m } |e_1| \Psi_{ M - m + 1 }  \rangle
= - \frac{ m }{ 2 ( m - \frac{1}{2} ) } \langle \Psi_{ M - m } |
 \Psi_{ M - m } 
\rangle =
\]

\beq
 - \frac{ m }{ 2 ( m - \frac{1}{2} ) } \langle \Psi_{ m } | \Psi_{ m } 
\rangle \;\;\; .
\label{res2}
\eneq

\noindent
Eq.(\ref{res2}) closes the induction relation:

\beq
\frac{
\langle \Psi_m | \Psi_m \rangle}{\langle \Psi_{ m-1} | \Psi_{ m - 1} \rangle}
 = \frac{ ( m - \frac{1}{2} ) ( M - m + 1 )}{
m ( M - m + \frac{1}{2} ) } 
\label{resos} \;\;\; .
\eneq
\noindent
The formula generated by recursion is:

\[
\langle \Psi_m | \Psi_m \rangle = \prod_{j=1 \ldots m} \frac{ ( j  - \frac{1}{2
} )( M - j + 1 )}{ j ( M - m + j+ \frac{1}{2} )} C_M 
\]

\beq
=
\frac{ \Gamma [ M + 1 ] \Gamma [ m + \frac{1}{2} ] \Gamma [ M - m + \frac{1}{2
} ] }{ \Gamma [ M + \frac{1}{2} ] \Gamma [ m + 1 ] \Gamma [ M-m+1 ] 
} C_M \;\;\; .
\label{finnor1}
\eneq
\noindent
The constant $C_M$ is expressed in terms of Wilson's integral as

\[
C_M = ( \frac{N}{2 \pi i })^M \oint \frac{ d z_1}{  z_1} \ldots \oint
\frac{ d z_M}{  z_M } \prod_{ i \neq j }^M ( 1 - \frac{ z_i
}{ z_j} )^2
\]

\beq
  = N^M \frac{ ( 2 M )!}{ 2^M} \;\;\; .
\label{kwil}
\eneq
\noindent
Eqs.(\ref{finnor1},\ref{kwil}) complete the proof.

Now we work out the formula for the two-spinon eigenstates. In this case
$M=N/2 - 1$ and Eq.(\ref{eqe5}) becomes:

\begin{displaymath}
{\cal H} e_1 \Phi \Psi_{GS} - e_1 {\cal H} \Phi \Psi_{GS} 
\end{displaymath}

\begin{equation}
=
\times \Psi_{GS} \biggl[ \sum_{j=1}^M z_j^2 \frac{\partial}{\partial z_j}
+ (M-\frac{1}{2}) e_1 \biggr] \Phi \; \; \; ,
\end{equation}
\noindent
where again we work with the Hamiltonian in scaled units. 
Eq.(\ref{comm2}) now becomes:

\[
\sum_{ j =1}^M z_j^2 \frac{ \partial}{ \partial z_j} \Phi_{ \alpha
\beta }  =
2 [ e_1 \Phi_{ \alpha \beta} ]   
\]

\beq
+
\left[ M z_\alpha - z_\alpha^2 \frac{ \partial}{ \partial z_\alpha }
+ M z_\beta - z_\beta^2 \frac{ \partial}{ \partial z_\beta} \right]
\Phi_{ \alpha \beta } \;\;\; ,
\eneq
\noindent
where the state $\Phi_{ \alpha \beta}$ has been defined in Eq.(\ref{2sp1}).
Hence, by letting $e_1$ act onto the two-spinon eigenstate $\Phi_{mn}$, we
obtain

\[
[ {\cal H} - \frac{E_{GS}}{ ( J  2 \pi^2) / N^2}  ] [ e_1 \Phi_{mn} ]
- ( \lambda_{mn} + M - \frac{3}{2} ) [e_1 \Phi_{mn} ] 
\]

\[
=
\sum_{ j =1}^M z_j^2 \frac{ \partial}{ \partial z_j} \Phi_{mn}  =
\sum_k a^{mn}_k \sum_{ \alpha,\beta =1}^N \frac{ ( z_\alpha^* )^m}{ N}
\frac{ (z_\beta^*)^n}{ N} \biggl[ 2 [ e_1 \Psi_{\alpha \beta} ] 
\]

\beq
+
 ( M z_\alpha - z_\alpha^2 \frac{ \partial}{ \partial z_\alpha} + 
M z_\beta - z_\beta^2 \frac{ \partial}{ \partial z_\beta} ) 
\Psi_{ \alpha \beta }  \biggr] \;\;\; ,
\eneq
\noindent
which implies

\[
 [ {\cal H} -\frac{E_{GS}}{ ( J  2 \pi^2) /N^2 } - 
( \lambda_{mn} + M + \frac{1}{2} ) ] e_1 \Phi_{ mn} ]  
\]

\[
=
\sum_k a_k^{mn} \{ [ M - m - k +1 ] \sum_j b_j^{m+k-1  , n-k} 
\Phi_{ m+k-1+j , n-k -j} 
\]

\beq
+ [ M - n +k + 1 ] \sum_j b_j^{m+k , n-k-1} \Phi_{ m+k +j , n-k-1-j} \}
\;\;\; .
\label{pain}
\eneq
\noindent
(See eqs.(\ref{fmn},\ref{thebs}) for the definition of the coefficient 
$a_k^{mn},b_k^{mn}$ and eq.(\ref{lambdas}) for the definition of 
$\lambda_{mn}$.

Let us, now, take the scalar product of both sides of Eq.(\ref{pain}) with
$\Phi_{ m-1 , n}$. The result can be recast in the form

\beq
\frac{ \langle \Phi_{m-1 , n} | e_1| \Phi_{mn}  \rangle }{\langle 
\Phi_{m-1 , n } | \Phi_{m-1 , n} \rangle } =
- \frac{ M - m +1}{ 2 ( M - m +\frac{1}{2}) } \;\;\; .
\label{tres1}
\eneq
\noindent
On the other hand, by taking the scalar product of $\Phi_{ m , n-1}$ with
both sides of Eq.(\ref{pain}), we obtain:

\[
( \lambda_{ m , n-1} - \lambda_{ mn} - M - \frac{1}{2} ) 
\langle \Phi_{ m , n-1} | e_1| \Phi_{ mn}  \rangle 
\]

\[
=
\sum_k a^{mn}_k \{ [ M - m - k + 1 ] 
\]

\[
\times \sum_j b_j^{m+k-1 , n-k} 
\langle \Phi_{ m , n-1} | \Phi_{ m + k -1 +j , n - k -j } \rangle 
\]

\[
+
[ M - n + k +1 ] \sum_j b_j^{ m+k , n-k -1} 
\]

\beq
\times
\langle \Phi_{ m , n-1} |
\Phi_{ m+k + j , n - k - j -1} \rangle \;\;\; ,
\eneq
\noindent
which implies the relation:

\[
\frac{ \langle \Phi_{ m , n-1} | e_1| \Phi_{ m n}  \rangle}{\langle 
\Phi_{ m , n-1} | \Phi_{ m , n-1 } \rangle}
\]

\beq
=
 - \frac{ ( M - n + \frac{3}{2} ) ( m - n + 1 )^2}{ 2 ( m - n + \frac{3}{2}
) ( m - n + \frac{1}{2} ) ( M - n + 1 )} \;\;\; .
\label{tspres2}
\eneq
\noindent
In order to complete the proof, we need two more identities, which can be 
proved in the same way we did for Eq.(\ref{resos}):

\[
\frac{ \langle \Phi_{ m - 1 , n } |  e_1| \Phi_{ mn}  \rangle}{
 \langle \Phi_{ m n } | \Phi_{ m n } \rangle} = 
\langle \Phi_{ M - n  , M - m }  |e_1| \Phi_{ M- n , M - m + 1 }  
\rangle 
\]

\beq
=
 - \frac{ ( m + \frac{1}{2} ) ( m - n)^2 }{ 2 ( m - n + \frac{1}{2} ) 
m ( m - n - \frac{1}{2} ) } 
\eneq

\[
\frac{\langle \Phi_{m , n-1} | e_1 |\Phi_{mn}  \rangle }{\langle \Phi_{ mn} 
| \Phi_{ mn} \rangle }
= 
\frac{ \langle \Phi_{ M - n , M - m } | e_1 |\Phi_{ M - n + 1 , M - m } 
\rangle}{\langle \Phi_{ mn} | \Phi_{ mn} 
\rangle }  
\]

\beq
=
- \frac{ n}{ 2 ( n - \frac{1}{2} )} \;\;\; .
\eneq

\noindent
Hence, the proof is given by the following induction relations:

 \[
\frac{ \langle \Phi_{ mn} | \Phi_{ mn} \rangle }{ \langle
\Phi_{ m , n-1} | \Phi_{ m , n-1} \rangle } 
\]

\beq
=
\frac{ ( n - \frac{1}{2} ) ( M - n + \frac{3}{2} ) ( m - n+ 1)^2}{
n ( M - n + 1 ) ( m - n+\frac{3}{2} ) ( m - n + \frac{1}{2} ) }
\label{recu1}
\eneq

\[
\frac{ \langle \Phi_{ mn} | \Phi_{ mn} \rangle }{\langle \Phi_{ m - 1 , n } |
 \Phi_{ m - 1 , n} \rangle } 
\]

\beq
=
\frac{ ( M - m + 1 ) ( m - n + \frac{1}{2} ) ( m - n - \frac{1}{2} ) m}{
( M -m + \frac{1}{2} ) ( m - n )^2 ( m + \frac{1}{2} ) } 
\;\;\; .
\label{recu2}
\eneq

\noindent
Eqs.(\ref{recu1},\ref{recu2}) imply

\[
\langle \Phi_{ m n } | \Phi_{ m n } \rangle 
\]

\[
=
C_M^{'} \frac{ \Gamma [ m - n + \frac{1}{2} ] \Gamma [ m - n + \frac{3}{
2} ] }{ \Gamma^2 [ m - n + 1 ] } 
\]

\beq
\times
\frac{ \Gamma [ m + 1 ] \Gamma [ M - m +
\frac{1}{2} ] }{ \Gamma [ m + \frac{3}{2} ] \Gamma [ M - m +1 ] }
\frac{ \Gamma [ n + \frac{1}{ 2} ] \Gamma [ M - n + 1 ] }{ \Gamma [ 
n + 1 ] \Gamma [ M - n + \frac{3}{2} ] } \;\;\; , 
\label{bigboy}
\eneq
\noindent
and the constant $C^{'}_M$ is now given by:

\beq
C^{'}_M = N^M \frac{ ( 2 M )!}{ 2^M } \frac{ M + \frac{1}{2} }{ \pi} \;\;\;  .
\eneq

\end{document}